\DocumentMetadata{lang=en-US,pdfversion=2.0}
\documentclass[aps,prd,onecolumn,superscriptaddress,nofootinbib,floatfix]{revtex4-2}

\usepackage{amsmath,amssymb,bm,mathtools}
\usepackage{graphicx}
\usepackage{booktabs}
\usepackage{array}
\usepackage{longtable,tabularx}
\usepackage{tabularray}
\usepackage{comment}
\UseTblrLibrary{booktabs}
\usepackage{ragged2e}
\usepackage{xspace}
\usepackage{etoolbox}
\usepackage{microtype}
\usepackage{placeins}
\usepackage{needspace}
\usepackage{hyperref}
\newcommand{\ReviewTitle}{Quantum Technologies: System-Level Performance and Validation Priorities}
\hypersetup{
  colorlinks=true,
  allcolors=blue,
  pdftitle={\ReviewTitle},
  pdfauthor={Slava G. Turyshev},
  pdfsubject={Physics-based review of quantum computing, simulation, communication, sensing, timing, random-number generation, and enabling technologies},
  pdfkeywords={quantum technologies, quantum computing, quantum error correction, quantum simulation, quantum communication, quantum sensing, optical clocks, quantum metrology, quantum random-number generation, quantum engineering},
  pdfdisplaydoctitle=true,
  bookmarksopen=true
}

\newcommand{\Tr}{\mathrm{Tr}}
\newcommand{\ket}[1]{\left|#1\right\rangle}

\newcommand{\braket}[2]{\left\langle#1|#2\right\rangle}
\newcommand{\kB}{k_{\mathrm B}}

\newcommand{\ii}{\mathrm i}

\newcommand{\NR}{\ensuremath{\mathrm{NR}}}

\AtBeginEnvironment{table}{\small}
\AtBeginEnvironment{table*}{\small}
\AtBeginEnvironment{tblr}{\small}
\AtBeginEnvironment{longtblr}{\small}
\AtBeginEnvironment{longtable}{\small}
\AtBeginEnvironment{thebibliography}{\footnotesize}

\begin{document}

\title{Quantum Technologies: System-Level Performance and Validation Priorities}

\author{Slava G. Turyshev}
\affiliation{Jet Propulsion Laboratory, California Institute of Technology,\\
4800 Oak Grove Drive, Pasadena, CA 91109-0899, USA}

\date{\today}

\begin{abstract}
Performance claims in quantum technology are not properties of
hardware alone. They are properties of a declared task, system
boundary, normalization denominator, uncertainty or security
convention, and comparator. We perform a cross-domain  analysis spanning quantum computing, simulation, communication, sensing, clocks, randomness generation, and their enabling technologies. The analysis changes substantive conclusions in several representative cases. For the longest direct finite-key quantum key distribution case analyzed here, the same final key gives rates differing by a factor of $1.91$ when normalized by complete acquisition rather than transmission-active time. In Advanced LIGO, $6.1\,\mathrm{dB}$ peak quantum-noise reduction coexists with a $0.534$ coincident analysis-ready fraction, separating detector-level gain from delivered observing service. In quantum computation and photonic sampling, matching the observable, error tolerance, loss model, sample count, amortization, and classical hardware moves or reverses published crossover claims. Across domains, the recurring limits are correlated error, multiplicative interface loss, thermal and nonequilibrium occupation, calibration covariance, measurement efficiency, fabrication yield, and control latency. A quantum advantage is therefore established only for a fixed task and boundary when the accepted output outperforms the best documented alternative at matched accuracy, elapsed time, availability, and lifecycle cost. The resulting framework identifies the measurements required to convert component records into reproducible system capability. Credible progress is defined by reproduced logical workloads, prospectively validated simulations, repeater links outperforming direct transmission, long-duration calibrated sensors and clocks,  integrated hardware with predictable yield and reliability.

\end{abstract}

\maketitle

%\tableofcontents

\section{Introduction}
\label{sec:introduction}

Quantum technologies exploit controlled preparation, coherent evolution, quantized transitions, nonclassical correlations, and quantum measurement to perform computation, simulation, communication, sensing, timing, and randomness generation. The field has moved beyond few-degree-of-freedom demonstrations: processors contain tens to thousands of controlled physical degrees of freedom, optical clocks resolve fractional frequency shifts below $10^{-18}$, quantum links operate through losses exceeding $100\,$dB, and squeezed states are incorporated into operating observatories. The scientific problem is to determine when the prepared quantum resource survives control, readout, interfaces, environmental coupling, calibration, packaging, and repeated operation sufficiently well to improve a complete function.

Quantum technologies are increasingly important for modern engineering because timing, inertial reference, secure communication, and precision measurement underpin navigation, critical-infrastructure protection, environmental observation, scientific instrumentation, and space missions. They can extend operation beyond conventional limits set by inertial drift, terrestrial-fiber attenuation, quantum measurement noise, and growth of many-body state spaces. Atom interferometers and optical clocks are being developed for gravimetry, geodesy, inertial navigation, and resilient timing; transportable gravimeters, mobile optical-clock ensembles, and a cold-atom gyroscope in orbit have already demonstrated operation beyond stationary laboratories \cite{Bongs2019,Menoret2018,Hilton2025,Li2025}. Satellite
entanglement distribution, satellite-to-ground quantum key distribution (QKD), and an integrated trusted-relay $4600\,\mathrm{km}$ space-to-ground network have established physical-layer capabilities for secure links, although loss, trust architecture, weather, pass duration, and availability remain system constraints \cite{Yin2017,Liao2017,Chen2021QKD}.
Frequency-dependent squeezed readout now improves operating gravitational-wave observatories, while clocks and quantum sensors support precision tests of fundamental physics, including searches for ultralight dark matter
\cite{Capote2025LIGOO4,Bongs2019,Ludlow2015,Kobayashi2022DM}. Programmable quantum simulation is experimentally established, but utility beyond strong classical methods and application-level fault-tolerant computing remain emerging and task dependent
\cite{Scholl2021,Georgescu2014,Preskill2018}. Realizing their broader value requires scalable control, low-loss interfaces, reproducible fabrication, calibration traceability, reliability, and sustained deployment---the constraints examined below.

The physical origin of the component-to-system gap is domain dependent. Physical-qubit count does not determine logical error or executable circuit depth; coherence time must be compared with the complete gate--measurement--reset cycle; and component fidelity does not predict a workload subject to leakage, correlated faults, and drift. Sensor sensitivity is inseparable from bandwidth, dynamic range, bias, dead time, and calibration. Clock instability, systematic uncertainty, transfer error, and holdover describe different processes. Point-to-point key rate does not determine network availability, and source brightness or conditional fidelity does not determine the entanglement rate after collection, conversion, switching, storage, and detection. The review therefore asks not only which physical mechanism sets the governing relation, but also which task, system boundary, denominator, uncertainty or security convention, and comparator make the resulting performance claim well defined.

Domain-specific reviews provide deeper coverage of individual
platforms, devices, and physical mechanisms. The contribution here is orthogonal: performance claims from computing, communication, sensing, and metrology are reconstructed only after fixing the task, system boundary, denominator, uncertainty or security convention, and comparator. Here ``denominator'' includes both literal normalizations and accepted-event or resource boundaries: per physical operation or logical cycle; per emitted pulse, detected event, active second, or elapsed second; per incident or absorbed photon; per calibrated observing interval; or per accepted workload result. A headline fidelity, rate, sensitivity, or runtime is therefore not treated as a portable property of the hardware.

Fixing these denominators changes substantive conclusions in four
cases developed below. For the IBM Eagle kicked-Ising experiment,
holding the observable and absolute-error target fixed allowed later
classical methods to reproduce the reported quantities with error
below $0.01$, reversing the specific conclusion of practical classical
inaccessibility without implying that arbitrary 127-qubit circuits are
easy \cite{Kim2023Utility,Tindall2024Eagle,Begusic2024}.
For photonic Gaussian boson sampling, replacing an exact ideal-sampler comparison with a loss-aware device model evaluated against the published validation observables removed the original crossover for Borealis and Jiuzhang~3.0
\cite{Madsen2022Borealis,Deng2023Jiuzhang3,Oh2024GBSMPS}.
For the $421.1\,\mathrm{km}$ finite-key QKD experiment, the same
$22\,124$ secret bits yield $0.2540\,\mathrm{bit\,s^{-1}}$ over the
complete acquisition interval and $0.484\,\mathrm{bit\,s^{-1}}$ over transmission-active time, a factor of $1.91$ produced solely by the time denominator \cite{Boaron2018}. For Advanced LIGO, quantum-noise reduction up to $6.1\,\mathrm{dB}$ and an O4a coincident analysis-ready fraction of $0.53397$ refer to different denominators: one is a detector-level gain in selected frequency bands, whereas the other is delivered calibrated observing time
\cite{Capote2025LIGOO4,GWOSCO4a2025}.
These are not corrections to the underlying experiments. They show
that the scientific or engineering conclusion can change when the
denominator and system boundary are made explicit.

The source base was assembled by targeted searches of Crossref, arXiv, Web of Science and Scopus records available to the author, and the American Physical Society (APS), Nature Portfolio, Science and the American Association for the Advancement of Science (AAAS), the Institute of Electrical and Electronics Engineers (IEEE), the Association for Computing Machinery (ACM), the National Institute of Standards and Technology (NIST), the International Bureau of Weights and Measures (BIPM), the European Telecommunications Standards Institute, and institutional repositories. Foundational papers from 1950 onward and technical results available through 14 August 2026 were considered. Primary experimental articles were included when the configuration, measurand, operating conditions, units, and uncertainty or security convention could be recovered; reviews and standards supplied context and definitions. Preprints were used only when no version of record was available and are identified as such; publisher corrections were applied, retracted work was excluded, and targeted citation searches included classical challenges, negative comparisons, replication studies, and population-level manufacturing results. Best-reported values were retained only with their operating point, while representative values were selected to cover each major platform, function, and limiting mechanism. The resulting source base contains approximately one hundred primary experiments, sufficient for a mechanism-resolved technical synthesis but not for estimating field-wide prevalence or typicality.

This review does not define a universal quantum-technology score.
Instead, it asks which performance claims remain valid after the
task, system boundary, denominator, convention, and comparator have been fixed. Comparisons are therefore made only within common functions, using compatible definitions and retaining the operating configuration, accepted-event boundary, uncertainty or security convention, and classical reference. Numerical values remain attached to the cited configuration and are not combined across apparatuses without an explicit physical model. In the text and tables, not reported (\NR{}) denotes a quantity absent from the cited source or its accessible supplement; it is not interpreted as zero or as not applicable. For communication systems, quantum bit error rate (QBER) denotes the fraction of sifted key bits or measurement outcomes that disagree under the stated protocol. Table~\ref{tab:field_map} maps the major domains without ranking them. Sections~\ref{sec:computing}--\ref{sec:qrng} treat computing, simulation, communication, sensing, clocks, and randomness generation; Sec.~\ref{sec:enabling} treats materials, fabrication, photonics, cryogenics, electronics, packaging, calibration, and demonstrated AI-assisted engineering; and Sec.~\ref{sec:limits} synthesizes the dominant mechanisms and experimentally testable paths forward. Appendix~\ref{app:abbreviations} provides recurring abbreviations used across the review.
  
\begin{table}[tbp]
\caption{Physics-based map of the principal quantum-technology domains. The rows describe distinct functions and are not a common performance scale.}
\label{tab:field_map}
\centering
\begin{tblr}{colspec={Q[l,wd={0.125\linewidth}] Q[l,wd={0.165\linewidth}] Q[l,wd={0.23\linewidth}] Q[l,wd={0.25\linewidth}] Q[l,wd={0.20\linewidth}]}, cells={valign=t}, rowsep=0.65pt, colsep=1.8pt}
\toprule
Domain & Resource and repre- sentative platforms & Output and governing metrics & Dominant limitations & Demonstrated status \\
\midrule
Computing and error correction & Coherent gates and entanglement; circuits, ions, atoms, photons, spins & Verified logical result; physical and logical error, code distance, depth, non-Clifford rate, wall time & Correlated error, leakage, routing, decoding, calibration, verification & Repeated logical memories; early logical operations \\
Quantum simulation & Engineered Hamiltonians; atoms, ions, gases, circuits, photons, molecules & Correlations, spectra, phase boundaries, dynamics; model and observable uncertainty & Model discrepancy, finite size, decoherence, postselection, verifiability & Quantitative observables in selected regimes \\
Communication and networking & Single photons, interference, entanglement, memories & Secret key, Bell pair, state transfer; loss, quantum bit error rate (QBER), fidelity, rate, memory time, availability & Exponential loss, background, phase control, interface efficiency, waiting-time tails & QKD fielded; satellite links demonstrated; repeaters experimental \\
Sensing and imaging & Phase, spin, squeezing, correlations, back-action & Calibrated field, motion, image or waveform; sensitivity, bandwidth, range, drift, duty factor & Technical noise, transfer error, environmental coupling, loss, dead time & Operational examples and laboratory or field prototypes \\
Clocks and metrology & Atomic, ionic, molecular, or nuclear transitions & Frequency, time, ratio, holdover; Allan deviation, systematic and transfer uncertainty & Local-oscillator noise, shifts, dead time, link covariance, autonomy & Atomic time scales operational; optical and nuclear systems developing \\
Quantum randomness & Measurement entropy from optical or electronic quantum noise & Extracted bits; min-entropy, rate, statistical distance, health-test failure & Source-model error, side information, digitizer and extractor implementation & High-rate laboratory and integrated implementations; selected entropy-source certifications; sustained service evidence limited \\
Enabling technologies & Low-loss generation, control, conversion, readout, packaging & Coherence, insertion loss, noise, heat load, yield, lifetime, calibration interval & Materials defects, process variation, cryogenic load, phase noise, packaging, control scaling & Foundry-compatible and population-scale components \\
\bottomrule
\end{tblr}
\end{table}

\subsection{Physical mechanisms shared across platforms}

For weak system--bath coupling, short environmental correlation time, and time-homogeneous completely positive trace-preserving dynamics, the generator of a Markovian quantum dynamical semigroup has the Gorini--Kossakowski--Sudarshan--Lindblad form \cite{Gorini1976,Lindblad1976}
\begin{equation}
\dot{\rho}=-\frac{\ii}{\hbar}[H,\rho]
+\sum_j\left(L_j\rho L_j^\dagger-\frac{1}{2}\{L_j^\dagger L_j,\rho\}\right).
\label{eq:lindblad}
\end{equation}
Here $H$ contains the controlled and static Hamiltonian terms and $L_j$ represent platform-specific channels: dielectric and quasiparticle loss in superconducting circuits, spontaneous emission and motional heating in ions, Rydberg decay and atom loss in neutral atoms, photon loss in optical systems, and spin relaxation or spectral diffusion in solid-state defects. For a two-level amplitude-damping rate $\Gamma_1$ and independent pure-dephasing rate $\Gamma_\phi$, the off-diagonal density-matrix element decays at $\Gamma_2=\Gamma_1/2+\Gamma_\phi$. Thus $T_2^{-1}=(2T_1)^{-1}+T_\phi^{-1}$ only for exponential, effectively Markovian channels. Quasistatic noise, $1/f$ spectra, pulse-dependent filtering, non-Markovian baths, and correlated faults require explicit noise spectra or stochastic process models. The experimentally relevant ratio is therefore not $T_2$ alone but the number and type of coherent, measurement, reset, and feedback operations that fit within the measured noise process.

Preparation and readout are represented by a quantum instrument $\{\mathcal I_x\}_x$, where each outcome map is completely positive and trace nonincreasing \cite{Kraus1983,NielsenChuang2010}:
\begin{equation}
\begin{aligned}
\mathcal I_x(\rho)&=\sum_{\alpha}K_{x\alpha}\rho K_{x\alpha}^{\dagger},
& E_x&=\sum_{\alpha}K_{x\alpha}^{\dagger}K_{x\alpha},\\
p(x|\rho)&=\Tr\!\left[\mathcal I_x(\rho)\right]=\Tr(E_x\rho),
& \rho_x&=\frac{\mathcal I_x(\rho)}{p(x|\rho)},\\
\sum_x\mathcal I_x&\ \text{is trace preserving},
& \sum_{x,\alpha}K_{x\alpha}^{\dagger}K_{x\alpha}&=I.
\end{aligned}
\label{eq:measurement}
\end{equation}
The operators $E_x$ form the associated positive operator-valued measure, and the normalized conditional state $\rho_x$ is defined when $p(x|\rho)>0$. The former single-Kraus-per-outcome expression is recovered as the special case $K_{x\alpha}=M_x\delta_{\alpha 1}$. Detector inefficiency changes $p(x|\rho)$; demolition, crosstalk, and measurement-induced transitions change $\rho_x$. These distinctions determine readout error in processors, dead time and contrast in clocks and sensors, heralding probability in photonic systems, and the error propagation of repeated syndrome extraction. No-cloning follows from preservation of inner products: a unitary satisfying $U\ket{\psi}\ket{0}=\ket{\psi}\ket{\psi}$ for all states would require $\braket{\phi}{\psi}=\braket{\phi}{\psi}^{2}$, which is possible only for identical or orthogonal states \cite{WoottersZurek1982,Dieks1982}. Quantum networks must therefore use measurement and repreparation, teleportation, entanglement swapping, or error-corrected transmission rather than transparent amplification of unknown states.

Serial interfaces are governed by probability conservation through a cascade. If interface $i$ transmits a photon, excitation, or successfully conditioned event with probability $\eta_i$, then $n_i=\eta_i n_{i-1}$ and
\begin{equation}
\eta_{\rm sys}=\prod_{i=1}^{M}\eta_i,
\qquad
\ln\eta_{\rm sys}=\sum_i\ln\eta_i.
\label{eq:serial_loss}
\end{equation}
For small fractional losses $\epsilon_i=1-\eta_i$, $\eta_{\rm sys}\simeq\exp(-\sum_i\epsilon_i)$; in decibels, losses add exactly. If the $\eta_i$ are independently calibrated, $(u_{\eta_{\rm sys}}/\eta_{\rm sys})^2\simeq\sum_i(u_{\eta_i}/\eta_i)^2$, whereas common coupling or alignment errors introduce covariance. Equation~\eqref{eq:serial_loss} controls photonic source-to-detector efficiency, microwave--optical conversion chains, remote-entanglement success, and detector-level squeezing. A hundred interfaces at $0.1\,$dB each transmit only $10\%$.

For a harmonic mode in a Gibbs state, summing the Bose--Einstein occupation over the ladder $E_n=hf(n+1/2)$ gives
\begin{equation}
\bar n_{\rm th}=\frac{1}{\exp(hf/\kB T)-1}.
\label{eq:thermal_occupation}
\end{equation}
The limits are $\bar n_{\rm th}\simeq\kB T/(hf)$ for $hf\ll\kB T$ and $\bar n_{\rm th}\simeq\exp[-hf/(\kB T)]$ for $hf\gg\kB T$. At $5\,$GHz, $hf/\kB=0.240\,$K, giving $\bar n_{\rm th}=6.2\times10^{-6}$ at $20\,$mK and $3.69$ at $1\,$K. These equilibrium values are lower bounds when infrared leakage, imperfect attenuation, local dissipation, quasiparticles, or hot interconnects populate the device mode.

Uncertainty propagation is equally scale dependent. For an inferred vector $\hat{\boldsymbol\theta}=\boldsymbol f(\boldsymbol v)$, first-order linearization about the operating point gives $C_\theta\simeq J C_vJ^{\mathsf T}$, with $J_{ij}=\partial f_i/\partial v_j$. For $M$ equal-variance terms $\sigma^2$ and common pairwise correlation $\rho$,
\begin{equation}
\frac{u_c}{u_{\rm ind}}=\sqrt{1+(M-1)\rho},
\qquad -\frac{1}{M-1}\leq\rho\leq1.
\label{eq:corr_inflation}
\end{equation}
This follows from $u_c^2=M\sigma^2+M(M-1)\rho\sigma^2$ and $u_{\rm ind}=\sqrt{M}\sigma$. For $M=100$, $\rho=0.01$ inflates the uncertainty by $1.41$ and $\rho=0.05$ by $2.44$. The relation is a diagnostic for clock-shift budgets, array calibration, common local-oscillator noise, and correlated readout; nonlinear or non-Gaussian models require Monte Carlo, Bayesian, or interval propagation.

\begin{figure}[tbp]
\centering
\includegraphics[alt={Four analytical panels showing phase-estimation scaling with small-N lossy asymptotes masked, squeezing loss, bosonic thermal occupation, and correlated-uncertainty inflation.},width=0.96\linewidth]{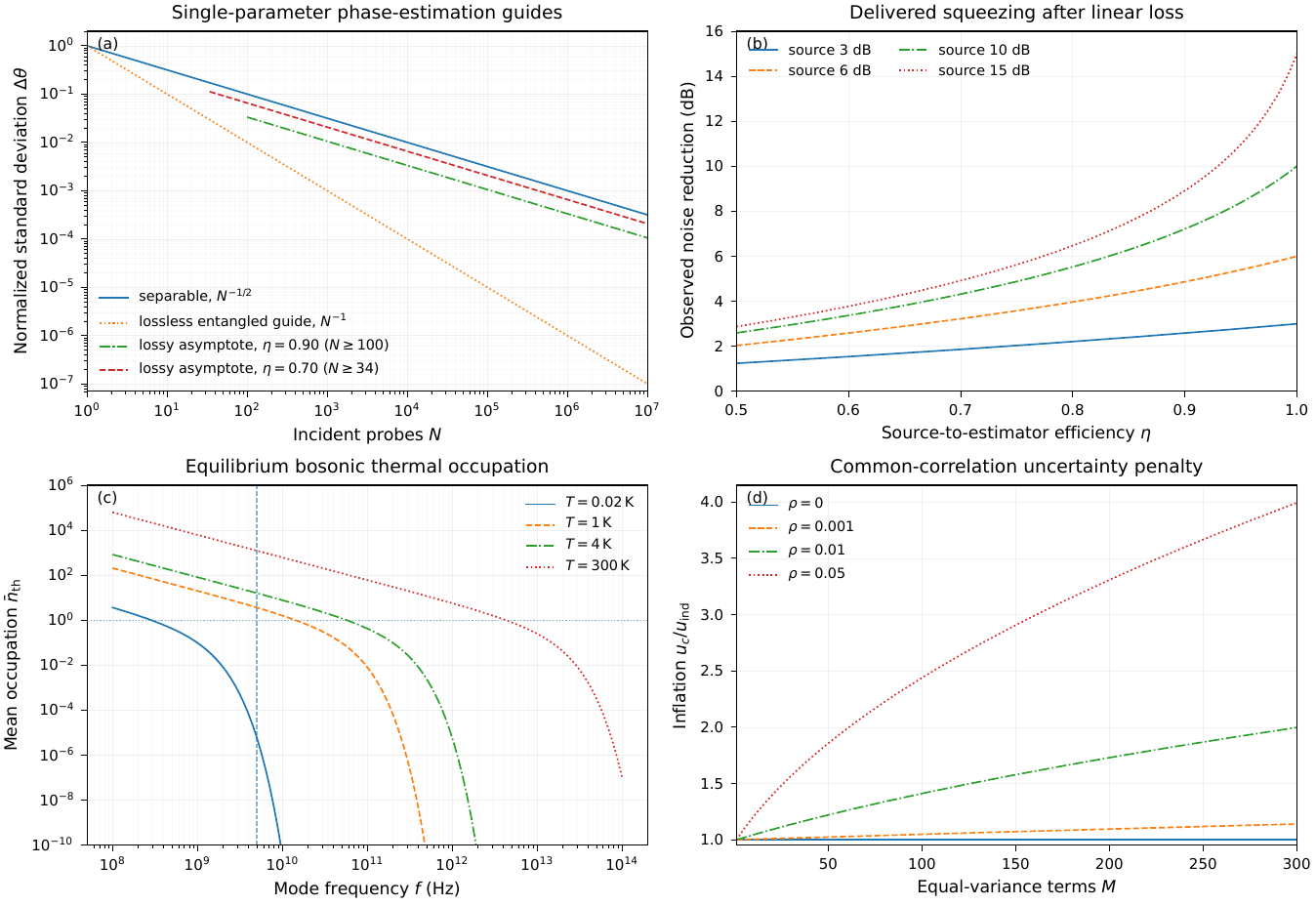}
\caption{Cross-cutting physical relations. (a) Local single-parameter guides $\Delta\theta=N^{-1/2}$ for separable probes, the lossless $N^{-1}$ guide, and the lossy channel-extension asymptote $\Delta\theta_{\rm er}=[(1-\eta)/(\eta N)]^{1/2}$ for $\eta=0.90$ and $0.70$ \cite{Demkowicz2012}. Here $N$ is the incident-probe count, the single-probe standard deviation is normalized to unity, and independent erasure occurs after encoding without heralding, multiplexing, correlated loss, or architecture-specific error correction. Because the expression is not finite-$N$ tight, each lossy curve is plotted only where $N(1-\eta)\geq10$: $N\geq100$ for $\eta=0.90$ and $N\geq34$ for $\eta=0.70$. The omitted small-$N$ region is intentionally not represented by the asymptotic relation. (b) Linear-loss squeezing from Eq.~\eqref{eq:squeezing_loss}: source levels $S_{\rm src}=3,6,10,15\,$dB correspond to $v_{\rm in}=10^{-S_{\rm src}/10}$ and are evaluated over $0.5\leq\eta\leq1$. (c) Equilibrium occupation from Eq.~\eqref{eq:thermal_occupation} for $T=0.02,1,4,300\,$K over $10^8\leq f/\mathrm{Hz}\leq10^{14}$; the $5\,$GHz and $\bar n_{\rm th}=1$ guides are visual references, and nonequilibrium occupation is omitted. (d) Equicorrelated uncertainty inflation from Eq.~\eqref{eq:corr_inflation} for $\rho=0,0.001,0.01,0.05$ and integer $1\leq M\leq300$. All panels are deterministic analytical evaluations with no fitted parameters or confidence intervals.}
\label{fig:physical_limits}
\end{figure}

For a parameter $\theta$ encoded in a state family $\rho_\theta$, the quantum Cram\'er--Rao bound is \cite{Helstrom1976,Holevo1982,Giovannetti2004,Giovannetti2011,Degen2017}
\begin{equation}
\operatorname{Var}(\hat\theta)\geq\frac{1}{\nu F_Q(\theta)},
\label{eq:qcrb}
\end{equation}
for $\nu$ independent repetitions and a locally unbiased estimator. Attainability depends on the measurement, nuisance parameters, finite-sample bias, and multiparameter compatibility. Separable probes give $F_Q\propto N$; ideal lossless entangled probes can give $F_Q\propto N^2$, but fixed uncorrelated loss restores asymptotic $N^{-1/2}$ standard-deviation scaling with a modified coefficient \cite{Demkowicz2012}. For independent single-probe erasure with survival probability $\eta$, the channel-extension result used in Fig.~\ref{fig:physical_limits}(a) is
\begin{equation*}
\Delta\theta_{\rm er}\geq\sqrt{\frac{1-\eta}{\eta N}}.
\end{equation*}
It assumes phase encoding before independent loss, a fixed incident-probe count $N$, local estimation, and no heralding, multiplexing, correlated loss, or architecture-specific error correction. The expression is a valid lower bound at finite $N$ but was derived to characterize the asymptotic loss-dominated regime and is generally not tight at small $N$. The cited analysis identifies a crossover scale $N_\times\sim(1-\eta)^{-1}$. To avoid presenting the asymptote as finite-$N$ performance, Fig.~\ref{fig:physical_limits}(a) displays it only under the conservative plotting criterion $N(1-\eta)\geq10$, giving $N\geq100$ for $\eta=0.90$ and $N\geq34$ for $\eta=0.70$; the masked region requires a finite-$N$ optimization. The bound therefore identifies a physical opportunity, while Eq.~\eqref{eq:serial_loss}, detection efficiency, estimator bias, and technical noise determine the measured gain.

Figure~\ref{fig:physical_limits} evaluates these relations. The panels share mechanisms, not a performance axis. For the numerical rendering, panel (a) uses $500$ logarithmically spaced values of $N$ on $1\leq N\leq10^7$, with the two lossy asymptotes masked below their stated thresholds; panel (b) uses $400$ linearly spaced values of $\eta$ on $0.5\leq\eta\leq1$; panel (c) uses $700$ logarithmically spaced values of $f$ on $10^8\leq f/\mathrm{Hz}\leq10^{14}$; and panel (d) uses every integer $1\leq M\leq300$. The curves are deterministic evaluations of the stated equations; no interpolation between source data, fit, stochastic sampling, or statistical uncertainty interval is used.

\section{Quantum computing and error correction}
\label{sec:computing}

Quantum computation requires coherent control over a Hilbert space large enough to implement an algorithm while suppressing accumulated physical, measurement, and control errors. The relevant physical question is not how many qubits are present, but how the error per operation, circuit connectivity, logical encoding, classical feedback, and resource-state production determine the probability and wall time of a correct result. Under weak Markovian noise, the ratio of coherence time to the complete gate--measurement--reset cycle provides only an upper bound on usable depth; leakage, coherent miscalibration, crosstalk, and nonstationary correlated faults can dominate well before that limit.

\subsection{Physical platforms and control architectures}

Superconducting circuits provide fast gates, lithographic coupling, dispersive readout, and mature microwave control \cite{Krantz2019,Kjaergaard2020,Blais2021}. Their scaling constraints include frequency crowding, leakage, cryogenic wiring, calibration drift, crosstalk, decoder throughput, and nonlocal error bursts. Controlled ionizing-radiation experiments increased nonequilibrium quasiparticle density and qubit relaxation, while lead shielding reduced the effect \cite{Vepsalainen2020Radiation,Vepsalainen2020Correction}. Correlated charge jumps extended beyond $600\,\mu$m and coincided with transient relaxation degradation across a millimeter-scale chip \cite{Wilen2021Correlated}. On a 26-qubit Sycamore subset, high-energy impacts increased simultaneous decay errors from a baseline near four to as many as 24, spread from a localized onset to chip-wide response on a roughly $1\,$ms scale, and recovered with a time constant near $25\,$ms \cite{McEwen2022Cosmic}. A $266.5\,$h synchronous measurement of ten transmons found all correlated events at $1/(101\pm1)\,$s and a cosmic-ray component at $1/(592^{+48}_{-41})\,$s, accounting for $17.1\pm1.3\%$ of events; cosmic-ray events most frequently affected all ten qubits, with device-dependent recovery constants near $0.7$ and $6\,$ms \cite{Harrington2025Cosmic}. In the cited transmon devices and observation intervals, these bursts produced transient spatial correlations and recovery dynamics that a purely local stationary fault model does not describe during the affected syndrome cycles. The measurements do not establish universal burst rates, spatial statistics, or recovery times for other layouts, shielding configurations, laboratories, or space environments; they instead motivate device- and environment-specific monitoring, conditional calibration and filtering, and nonstationary fault models. Backside normal-metal reservoirs have reduced pair-breaking phonon flux by more than $20\times$ and correlated quasiparticle poisoning by approximately two orders of magnitude, demonstrating that phonon downconversion, gap engineering, event tagging, and erasure-aware decoding are architecture-level mitigation strategies \cite{Iaia2022Phonon}. Trapped ions provide identical qubits, high connectivity, high-fidelity optical control, and transport in quantum charge-coupled device (QCCD) architectures; a 98-qubit apparatus reported mean single-qubit, two-qubit, and state-preparation-and-measurement (SPAM) infidelities of $2.5(1)\times10^{-5}$, $7.9(2)\times10^{-4}$, and $3.3(5)\times10^{-4}$, with parenthetical values reported as one-standard-deviation bootstrap uncertainties \cite{Ransford2026}. Gate duration, parallel zones, motional heating, optical delivery, and modular interconnects remain central.

Neutral atoms combine reconfigurable geometry, Rydberg interactions, long storage, and loss-resolved imaging. One physical-array experiment trapped more than 6100 atoms in approximately 12000 sites, with $T_2=12.6(1)\,$s, approximately 23-min lifetime, survival probability $99.98952(1)\%$, and imaging fidelity above $99.99\%$ in that apparatus \cite{Manetsch2025}. These storage and imaging quantities are not simultaneous logical-computation metrics. Semiconductor spin qubits offer dense electrostatic control and compatibility with 300-mm fabrication, while valley splitting, charge noise, tuning automation, long-range coupling, and cryogenic electronics limit system yield \cite{Loss1998,Hanson2007,Zwanenburg2013,Huang2024,Steinacker2025,Neyens2024}. Photonic architectures use multiplexed sources, low-loss circuits, interference, measurement, feed-forward, and photon-number-resolving detection. A foundry-compatible module reported conditional SPAM, Hong--Ou--Mandel (HOM), fusion, and interconnect fidelities above $99\%$, but unconditional source-to-detector probability and factory throughput remain separate quantities \cite{Alexander2025}.

Color-center computing architectures use optically addressable electron spins, nearby nuclear memories, and photonic links. Current evidence is strongest for network-node primitives and long-lived local registers rather than large monolithic processors; remote solid-state modules and memory-assisted entanglement therefore appear in the networking comparison rather than being assigned a processor-scale qubit count \cite{Stas2026}. Topological approaches seek hardware-level suppression through non-Abelian excitations or Majorana modes, but no cited experiment establishes a braiding-protected, application-level logical processor. Materials signatures, parity lifetimes, readout, control, and a fault-tolerant logical operation must be kept distinct from the theoretical protection mechanism.

Bosonic encodings, quantum low-density parity-check (LDPC) codes, and topological proposals change the overhead and connectivity problem rather than eliminate it \cite{Panteleev2022LDPC,Bravyi2024LDPC,Gottesman2001GKP,Mirrahimi2014Cat,Nayak2008,Lutchyn2010,Oreg2010}. Quantum annealers and coherent Ising machines target specialized optimization and sampling tasks; coherent dynamics have been demonstrated in a programmable 2000-qubit Ising chain \cite{King2022Annealing}, but comparative performance remains highly dependent on embedding, instance distribution, annealing schedule, precision, and the selected classical solver.

\subsection{Configuration-resolved hardware comparison}

Table~\ref{tab:computing} compares hardware configurations using quantities relevant to computation and error correction. The table is not a platform ranking: each row identifies a different apparatus and operating point. Population-level evidence exists for selected wafers and devices, but independent replication, long-duration duty factor, calibration burden, and packaged lifetime are rarely reported together with logical or application output.

\begin{table}[tbp]
\caption{Selected quantum-computing configurations organized by common functional categories. Quantum error correction (QEC) denotes logical encoding with syndrome-based error suppression or correction. Every entry refers to the cited operating point; the rows use different physical architectures and are not placed on a common scalar axis. Unless identified as analytical or independently recalculated, numerical entries are source-transcribed.}
\label{tab:computing}
\centering
\begin{tblr}{colspec={Q[l,wd={0.135\linewidth}] Q[l,wd={0.215\linewidth}] Q[l,wd={0.195\linewidth}] Q[l,wd={0.195\linewidth}] Q[l,wd={0.215\linewidth}]}, cells={valign=t}, rowsep=1.3pt, colsep=1.3pt}
\toprule
Platform / configuration & Error or accepted output & Scale, connectivity, and timing & Reproducibility and infrastructure evidence & Principal limiting mechanism and required advance \\
\midrule
Superconducting surface-code memory \cite{Acharya2025} & $p_L=(1.43\pm0.03)\times10^{-3}$ per cycle at $d=7$; $\Lambda=2.14\pm0.02$; separate $d=5$ real-time result & $101=49+48+4$ qubits (data, syndrome, boundary leakage-removal); $1.1\,\mu$s syndrome cycle in the real-time configuration & Repeated cycles and public methods in one processor family; wafer population data exist separately \cite{VanDamme2024} & Repeated logical gates, correlated-error control, routing, factories, cryogenic control, and verified workload \\
Trapped-ion QCCD \cite{Ransford2026} & Mean single-qubit, two-qubit, and SPAM infidelities $2.5(1)\times10^{-5}$, $7.9(2)\times10^{-4}$, and $3.3(5)\times10^{-4}$ & 98 $^{137}$Ba$^+$ qubits; transport-mediated all-to-all connectivity & Qubit- and gate-population statistics within one apparatus; optical and vacuum infrastructure not represented by the fidelity alone & Parallel zones, transport and gate time, motional heating, optical delivery, modular links, and logical error correction \\
Neutral atoms \cite{Manetsch2025,Bluvstein2026FT} & Physical storage and imaging above $6100$ atoms in one apparatus; separate $d=5$ four-round QEC error $0.62(3)\%$ per round & Reconfigurable tweezer arrays; atom-loss information used in the QEC characterization & Large-array storage and bounded QEC demonstrated in different configurations; no combined processor specification & Many-round gate-interleaved logical operation, loss management, fast feed-forward, decoder scaling, and duty factor \\
Photonic module \cite{Alexander2025} & Conditional SPAM $99.98(1)\%$, HOM $99.50(25)\%$, fusion $99.22(12)\%$, interconnect $99.72(4)\%$ & Foundry-compatible silicon photonics with cryogenic detection & Process-compatible optical components and detector population; metrics are conditional on detection & Unconditional source-to-detector probability, multiplexing, switching, feed-forward, factory throughput, and packaging \\
Silicon spin devices \cite{Neyens2024,Steinacker2025} & Source-reported operations above $99\%$ in selected unit cells; $223/232$ twelve-dot arrays passed the full-device criterion (reported as $96\%$) & Dense electrostatic control in 300-mm fabrication & Device-population yield and threshold-voltage distributions; coupled logical arrays not demonstrated & Automated tuning, exchange and valley uniformity, cryogenic electronics, long-range coupling, and logical yield \\
\bottomrule
\end{tblr}
\end{table}

\subsection{Models, workloads, and evidence of useful computation}

A gate-model processor applies a sequence $U_m\cdots U_1$ and samples from $p(x)=|\langle x|U_m\cdots U_1|\psi_{\rm in}\rangle|^2$. Algorithms for factoring, unstructured search, phase estimation, amplitude estimation, Hamiltonian simulation, and quantum signal processing exploit structured interference \cite{Shor1997,Grover1997,LowChuang2019}. The accepted output is not a circuit execution alone, but a numerical result or sample distribution with a target error, confidence or failure probability, and verification procedure.

For chemistry and materials, relevant resources include logical qubits, logical depth, non-Clifford count and throughput, state preparation, measurement, classical preprocessing, and wall time \cite{Bauer2020,McArdle2020}. For sampling or optimization, the instance ensemble, objective, approximation ratio, success probability, and classical baseline must be specified. A quantum-advantage claim requires total cost per accepted result against the most capable documented classical algorithm and hardware for the same input and accuracy. The 127-qubit kicked-Ising experiment and its classical reassessment are analyzed in Sec.~\ref{sec:classical_challenges}.

A fault-tolerant workload specification must report logical width $N_L$ and depth $D_L$, non-Clifford count $N_T$ and depth $D_T$, logical-cycle and decoder times $\tau_{\rm cyc}$ and $\tau_{\rm dec}$, verified magic-state production rate $R_T$, measurement requirement $N_{\rm meas}$, workload failure probability $P_{\rm fail}$, and wall time $t_{\rm wall}$. Before compilation, queueing, input/output, and verification overhead, the physical lower bound is $t_{\rm wall}\gtrsim\max(D_L\tau_{\rm cyc},N_T/R_T)$; routing, feed-forward, factory contention, and recovery increase it.

\subsection{Classical challenges and moving quantum--classical crossovers}
\label{sec:classical_challenges}

Quantum advantage is not a static property of a processor. It is a comparison between a fixed quantum task and the best available classical algorithm, hardware, accuracy target, and cost boundary. Random-circuit sampling and the 127-qubit kicked-Ising experiment provide two well-documented sequences.

The 53-qubit, depth-20 random-circuit experiment produced $10^6$ samples in approximately $200\,$s; the original estimate assigned approximately $10^4$ years to the then-assumed Summit implementation \cite{Arute2019}. Improved contraction order, secondary storage, and distributed tensor processing rapidly changed that estimate. A full-amplitude study showed that the 53--54-qubit circuit family could be simulated on Summit on a time scale of days, at the cost of a different and storage-intensive output strategy \cite{Pednault2019}. A Sunway tensor-network implementation reported a $304\,$s sampling simulation using $41.9$ million cores and peak mixed-precision performance of $4.4\,$Eflop s$^{-1}$ \cite{Liu2021Sunway}. A later graphics processing unit (GPU) tensor-network calculation generated $10^6$ uncorrelated samples for the 53-qubit, 20-cycle Sycamore circuits at approximate fidelity $F\simeq0.0037$ in $15\,$h on 512 GPUs \cite{Pan2022Sycamore}. Its tensor network was contracted once and the cost was amortized over the sample set, whereas the secondary-storage study targeted all amplitudes and the Sunway result used a different parallel simulation objective. Contraction-order search, preprocessing, storage input/output, and sample amortization therefore belong to the task definition. These results do not alter the $200\,$s hardware execution; they move the crossover by changing the algorithm, memory hierarchy, classical hardware, and sometimes the output or fidelity target.

A more direct revision occurred for the IBM Eagle kicked-Ising experiment. The 127-qubit circuits contained up to 60 controlled-NOT (CNOT) layers and were compared with matrix-product-state (MPS) and isometric tensor-network state (isoTNS) calculations that failed in strongly entangling regimes \cite{Kim2023Utility}. Geometry-aware belief-propagation tensor networks subsequently reproduced the observables with greater precision \cite{Tindall2024Eagle}. Sparse-Pauli dynamics and mixed Schr\"odinger--Heisenberg tensor networks converged the same observables to absolute error below $0.01$, with many shallow points taking minutes to hours on central processing units or a consumer GPU and the reported calculations running orders of magnitude faster than the experiment \cite{Begusic2024}. This is a reversal of the specific inference that those observables lay beyond practical classical approximation; it is not a claim that arbitrary 127-qubit circuits are easy. The heavy-hex geometry, near-Clifford rotations, local observables, and limited operator entanglement created exploitable structure.

\paragraph{Photonic Gaussian boson sampling.}
Gaussian boson sampling (GBS) injects single-mode squeezed vacuum states into a passive interferometer $U$ and samples an output occupation vector $\boldsymbol n=(n_1,\ldots,n_m)$. For a pure, zero-displacement, lossless instance with $B=U\operatorname{diag}(\tanh r_j)U^{\mathsf T}$, photon-number-resolving (PNR) detection gives
\begin{equation}
p(\boldsymbol n)=\frac{\left|\operatorname{Haf}(B_{\boldsymbol n})\right|^2}{\boldsymbol n!\prod_j\cosh r_j},
\label{eq:gbs_probability}
\end{equation}
where $B_{\boldsymbol n}$ repeats rows and columns according to $\boldsymbol n$ and $\boldsymbol n!=\prod_i n_i!$ \cite{Hamilton2017GBS}. Threshold detection is instead governed by a Torontonian probability and has different collision constraints \cite{Quesada2018Threshold}. Complexity arguments rely on approximate Hafnian sampling and anticoncentration, while exact classical cost grows primarily with detected photon or click number rather than nominal mode count \cite{Quesada2020ExactGBS,Bulmer2022GBSBoundary,Ehrenberg2025GBS}. Total transmission, partial distinguishability, mode mismatch, phase error, collisions, detector resolution, and dark counts jointly define the experimental distribution. Sufficient loss and detector noise generate classically simulable regimes whose boundary depends on squeezing, transmission, and detector quality \cite{Qi2020NoisyGBS}.

Borealis implemented 216 squeezed time-bin modes with three-dimensional programmable connectivity and PNR detection. A 216-mode sample required $36\,\mu$s; the output had mean detected photon number 125 and events up to 219 photons, while the original exact-sampling estimate exceeded $9000$ yr per sample \cite{Madsen2022Borealis}. Jiuzhang 3.0 used pseudo-PNR detection, included partial distinguishability in its device model, registered up to 255 clicks, and produced one sample in $1.27\,\mu$s; the source estimated approximately $600$ yr on Frontier for a representative exact ideal sample and $3.1\times10^{10}$ yr for its hardest event \cite{Deng2023Jiuzhang3}. A later loss-aware MPS sampler exploited the reduced entanglement of the lossy Gaussian state. For its most demanding reported case, MPS construction required approximately $10\,$min and generation of $10^7$ samples approximately $62\,$min on a mode-parallel NVIDIA A100 GPU implementation; the resulting samples matched the ground-truth benchmarks at least as well as the experimental data \cite{Oh2024GBSMPS}. Consequently, the original Borealis and Jiuzhang 3.0 comparisons with exact samplers no longer establish a crossover against the strongest published loss-aware method under those validation observables. Jiuzhang 4.0 subsequently used 1024 squeezed states in 8176 modes, with $92\%$ source efficiency, $51\%$ overall efficiency, and events up to 3050 photons, and reported a renewed separation from then-current MPS methods \cite{Liu2026Jiuzhang4}; low-entanglement Gaussian-to-MPS constructions improve the tractable regime but identify the largest high-entanglement configurations as outside their demonstrated range \cite{Liu2026LowEntanglementGBS}.

Validation determines what remains of the advantage claim. Borealis evaluated exact three- to six-photon distributions in 16 modes, obtaining fidelity above $99\%$ and total-variation distance no larger than $6.5\%$, and used cross-entropy, Bayesian likelihood ratios, and low-order cumulants at larger scale \cite{Madsen2022Borealis}. Jiuzhang 3.0 used Bayesian tests and correlation functions against models including partial distinguishability and available spoofers \cite{Deng2023Jiuzhang3}. These statistics reject specified thermal, coherent, distinguishable-photon, squashed-state, mean-field, or low-order alternatives; they do not prove closeness to the complete target distribution. Cross-entropy can be optimized by a classical spoofer without sampling the target \cite{Oh2023Spoofing}, and low-order moments do not constrain all higher-order correlations. A persuasive photonic crossover therefore requires exact tests in tractable sectors, multiple independently chosen higher-order statistics, released samples and device models, and comparison with the strongest current classical sampler at the same loss, collision statistics, detector model, sample count, and acceptance threshold. The surviving claim is configuration specific: photonic processors generate exceptionally large nonclassical samples rapidly, but computational advantage must be re-established whenever a stronger classical model reproduces the validation suite.

Table~\ref{tab:classical_challenges} consolidates the three crossover histories. Its central conclusion is that a surviving advantage claim is conditional on a fixed output distribution or observable, accuracy and validation target, loss and noise model, sample count, and classical hardware; changing those quantities can move or reverse the crossover without changing the quantum experiment.

\begin{table}[tbp]
\caption{Representative revisions of quantum--classical crossover estimates. Runtime and accuracy are retained only when defined by the source; energy was not reported on a common boundary.}
\label{tab:classical_challenges}
\centering
\begin{tblr}{colspec={Q[l,wd={0.105\linewidth}] Q[l,wd={0.215\linewidth}] Q[l,wd={0.195\linewidth}] Q[l,wd={0.225\linewidth}] Q[l,wd={0.215\linewidth}]}, cells={valign=t}, rowsep=1.1pt, colsep=1.2pt}
\toprule
Quantum task & Original hardware and claim & Later classical method and hardware & Quantitative outcome & Interpretation \\
\midrule
Random-circuit sampling \cite{Arute2019} & 53 qubits, 20 cycles, $10^6$ samples in $\sim200\,$s; original classical estimate $\sim10^4$ yr & Secondary-storage simulation \cite{Pednault2019}; Sunway tensor contraction \cite{Liu2021Sunway}; sampling on 512 GPUs \cite{Pan2022Sycamore} & Days for full amplitudes; $304\,$s Sunway result under its sampling implementation; $15\,$h for $10^6$ samples at $F\simeq0.0037$ on 512 GPUs & Crossover reduced by new algorithms, memory and hardware; output definition and target fidelity must be matched before declaring a reversal \\
Kicked-Ising observables \cite{Kim2023Utility} & 127 qubits, up to 60 CNOT layers; mitigated observables compared with limited MPS/isoTNS calculations & Belief-propagation tensor networks \cite{Tindall2024Eagle}; sparse-Pauli and mixed-picture tensor networks \cite{Begusic2024} & Converged expectation values with absolute error $<0.01$; reported runtime below the quantum experiment for the studied observables & Direct reversal of the claimed classical inaccessibility for this model and observable set; not a result for generic circuits \\
Photonic GBS \cite{Madsen2022Borealis,Deng2023Jiuzhang3} & Borealis: 216 squeezed modes, $36\,\mu$s per sample, up to 219 photons; Jiuzhang 3.0: $1.27\,\mu$s per sample, up to 255 clicks; original exact estimates $>9000$ yr and $\sim600$ yr per representative sample & Loss-aware MPS sampling \cite{Oh2024GBSMPS}; cross-entropy spoofing analysis \cite{Oh2023Spoofing} & Approximately $10\,$min MPS construction plus $62\,$min for $10^7$ samples in the most demanding reported MPS case; classical samples reproduced published benchmarks at least as well as experiment & The original exact-sampler comparison was undermined by the later lossy-device model; Jiuzhang 4.0 is a newer source-reported crossover requiring independent reassessment \\
\bottomrule
\end{tblr}
\end{table}

The physics is the computational structure: tensor-network cost grows with contraction width and entanglement rather than qubit count alone; sparse-Pauli methods exploit low non-Clifford content and operator sparsity; local observables can have much smaller causal cones than a global state. A robust advantage experiment must therefore publish circuits and data, specify the accuracy and sample fidelity, and update the classical baseline through the final analysis.

\subsection{NISQ computation, mitigation, detection, and fault tolerance}

The noisy intermediate-scale quantum (NISQ) regime comprises processors large enough to execute classically nontrivial circuits but without scalable error correction \cite{Preskill2018}. Error mitigation reduces estimator bias by additional circuit executions, noise scaling, probabilistic cancellation, symmetry constraints, or learned corrections; it does not create an encoded logical subspace and its sampling cost can grow rapidly with circuit volume and noise \cite{Temme2017,Cai2023QEM}. Error detection discards or flags outcomes inconsistent with a code or symmetry, improving conditional fidelity at the cost of acceptance probability. These approaches must report both the corrected estimator and the total quantum--classical execution cost.

``Break-even'' error correction is task specific. A logical memory or operation is beyond break-even only when it outperforms the declared physical comparator under the same duration, initialization, measurement, and failure definition. Logical advantage is stronger: an encoded implementation must improve an application-relevant output, not merely one memory channel. Fault-tolerant quantum computing requires a complete set of protected preparations, measurements, logical gates, routing, non-Clifford resources, decoding, and recovery whose error can be reduced systematically below the application requirement with scalable overhead. Physical-qubit count alone does not locate a system on this progression.

\subsection{Error correction, real-time decoding, and logical operation}

For a rotated surface-code memory patch,
\begin{equation}
N_{\rm patch}\simeq2d^2-1
\label{eq:patch}
\end{equation}
counts data and syndrome qubits before routing, leakage removal, logical-gate ancillas, magic-state factories, input/output, spares, and classical control \cite{Fowler2012,Terhal2015}. At $d=7$, Eq.~\eqref{eq:patch} gives $97$ data-plus-syndrome qubits. The reported $101$-qubit implementation used $49$ data qubits, $48$ syndrome-measurement qubits, and four additional boundary leakage-removal qubits for data-qubit leakage removal \cite{Acharya2025}; the four-qubit difference is therefore an implementation overhead, not a discrepancy in the idealized patch count. A common below-threshold phenomenological guide is
\begin{equation}
p_{\rm L}(d)\simeq A\left(\frac{p}{p_{\rm th}}\right)^{(d+1)/2},
\label{eq:surface}
\end{equation}
where $A$, $p_{\rm th}$, and effective $p$ depend on circuit, schedule, leakage, correlations, decoder, and failure definition. It is not a universal processor model.

A superconducting distance-7 memory used 101 physical qubits and reported $(1.43\pm0.03)\times10^{-3}$ logical error per cycle, with a distance-suppression factor $\Lambda=2.14\pm0.02$ \cite{Acharya2025}. A separate distance-5 real-time pipeline reported $(3.5\pm0.1)\times10^{-3}$ per cycle, a $1.1\,\mu$s syndrome cycle, and $63\pm17\,\mu$s mean decoder latency while streaming up to $10^6$ cycles. The two operating points used different decoder pipelines and are not fitted here as a two-point scaling law. Subsequent work has continued to improve logical error through correction and detection and to demonstrate additional logical-operation primitives \cite{Paetznick2026,Butt2026}.

The neutral-atom fault-tolerance study reported $0.62(3)\%$ error per round for a distance-5 four-round characterization circuit and a $d=3$-to-$d=5$ ratio $2.14(13)$ using atom-loss information and a hybrid decoder \cite{Bluvstein2026FT}. This number is not the same quantity as the superconducting $\Lambda$: one is a bounded four-round ratio using erasure information, the other a fitted per-distance-step memory-scaling factor. The numerical coincidence is a useful warning against comparing labels without protocols.

\begin{figure}[tbp]
\centering
\includegraphics[alt={Measured superconducting surface-code logical-memory points, phenomenological below-threshold scenarios, and conditional memory-patch qubit floors.},width=0.96\linewidth]{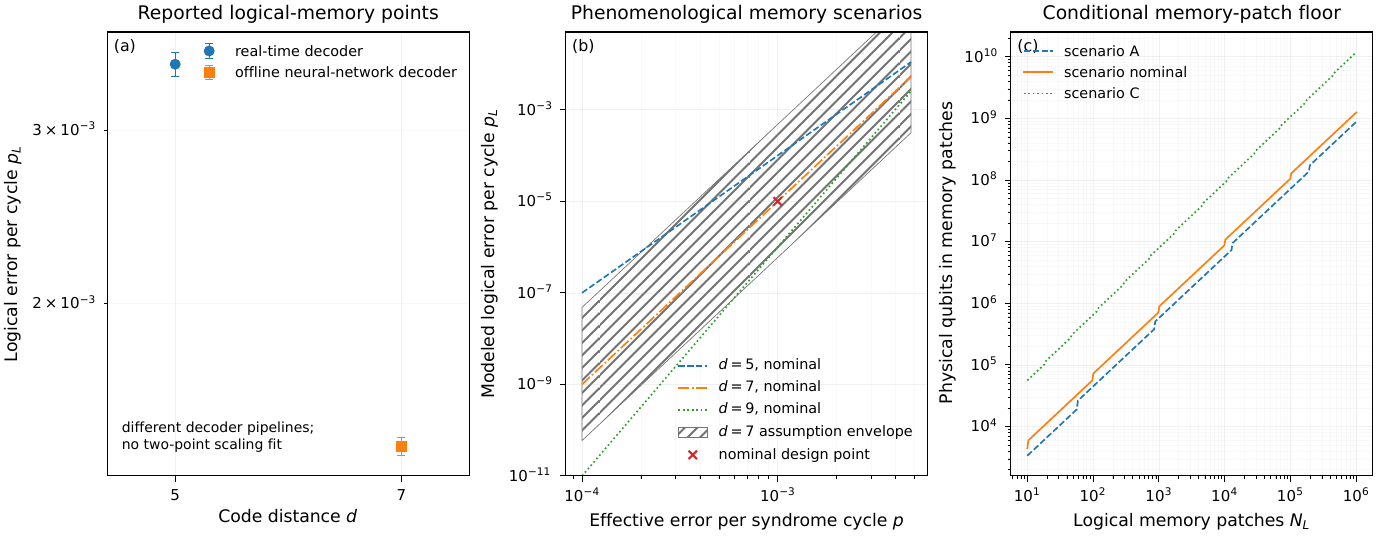}
\caption{Measured surface-code memory performance and model-conditioned scenarios. (a) Source-reported logical error per cycle $p_L=(3.5\pm0.1)\times10^{-3}$ at $d=5$ with the real-time decoder and $(1.43\pm0.03)\times10^{-3}$ at $d=7$ with the offline neural-network decoder; error bars are the reported one-standard-deviation uncertainties, and no two-point scaling fit is made \cite{Acharya2025}. (b) Equation~\eqref{eq:surface} with nominal $A=0.10$, $p_{\rm th}=10^{-2}$, $d=5,7,9$, and $10^{-4}\leq p\leq4.8\times10^{-3}$. The $d=7$ hatched assumption envelope is the pointwise minimum and maximum of the four corner curves $A\in\{0.03,0.30\}$ and $p_{\rm th}\in\{0.005,0.015\}$, evaluated on $480$ logarithmically spaced $p$ values and masked at $p\geq p_{\rm th}$; it is not a confidence interval. (c) Patch-only physical-qubit floors $N_L(2d^2-1)$ for $N_{\rm cyc}=10^6$, $320$ logarithmically spaced values on $10\leq N_L\leq10^6$, and scenarios $(A,p_{\rm th},p,\epsilon_{\rm app})$: A $(0.03,0.015,10^{-3},10^{-2})$, nominal $(0.10,0.010,10^{-3},10^{-2})$, and C $(0.30,0.005,2\times10^{-3},10^{-4})$. The minimum positive odd $d$ satisfying $p_L\leq\epsilon_{\rm app}/(N_LN_{\rm cyc})$ is used; $A$ is dimensionless, and $p_{\rm th}$, $p$, and $\epsilon_{\rm app}$ are probabilities. Panels (b), (c) are deterministic sensitivity calculations with no sampling, fit, or statistical interval; they omit logical gates, routing, non-Clifford-state factories, leakage, correlated noise, decoder failure, classical-control constraints, and verification and are not processor forecasts.}
\label{fig:fault}
\end{figure}

For $N_{\rm loc}$ logical fault locations and application failure allowance $\epsilon_{\rm app}$,
\begin{equation}
p_{\rm L}\lesssim\frac{\epsilon_{\rm app}}{N_{\rm loc}}
\label{eq:workload_target}
\end{equation}
provides a first design target. More generally, let $A_{ij}$ denote failure of the $j$th location of class $i$, and let $p_i=P(A_{ij})$ be the marginal failure probability in the implemented workload context. Then
\[
P_{\rm fail}=P\Big(\bigcup_{i,j}A_{ij}\Big)
\leq \sum_i N_i p_i
\]
is the union bound and does not require independent events. What must be validated are the probabilities inserted into the bound: an isolated-operation estimate need not equal the marginal probability under simultaneous control, common-mode disturbances, conditional error mechanisms, decoder state, or calibration drift. The approximation $P_{\rm fail}\simeq\sum_iN_ip_i$ additionally requires rare events with limited overlap; otherwise a context-matched joint model or direct workload measurement is required. Logical gates, factories, routing, leakage, decoder failures, and verification usually dominate over the memory-only floor. Figure~\ref{fig:fault} keeps measured logical-memory points separate from model-conditioned scenarios; its caption gives the complete deterministic parameter sets, grids, and decision rule.

%\FloatBarrier

Machine-learning methods enter this stack at the calibration, pulse-control, readout, and decoder layers rather than changing the computational model itself. Their contribution must therefore be reported as a change in a quantum-system metric---for example a same-device gate error, logical error, accepted-sample probability, calibration interval, or decoder latency---under the same circuit and operating point. AI-assisted decoding and calibration are treated below as classical control layers; their value is measured by logical error, latency, and robustness under the same circuit and noise process.

\subsection{Classical and machine-learning decoders}

QEC requires a decoder that maps a noisy detector history to a logical correction or frame update. Minimum-weight perfect matching (MWPM), union--find, belief propagation, tensor-network contraction, maximum-likelihood decoding (MLD), and learned decoders occupy different accuracy--latency--model-dependence regimes. Threshold values cannot be compared without matching code geometry, syndrome circuit, leakage and erasure treatment, measurement noise, correlations, and decoder implementation. For example, circuit-level surface-code studies using matching report thresholds from $0.502(1)\%$ to $1.140(1)\%$ depending on those choices \cite{Stephens2014}; union--find achieves almost-linear worst-case complexity $O[n\alpha(n)]$ and reported toric-code thresholds of $9.9\%$ with perfect syndrome and $2.6\%$ with faulty measurements, which are different noise models \cite{Delfosse2021}. Tensor-network MLD is exact in $O(n^2)$ for a restricted independent-noise, noiseless-syndrome setting and approximate in $O(n\chi^3)$ for more general models, with bond dimension $\chi$ controlling the accuracy--cost trade \cite{Bravyi2014Decoder}. Belief propagation with ordered-statistics decoding (BP+OSD) has reported a $9.9\pm0.2\%$ toric-code threshold for the paper's perfect-syndrome data-noise model, while generalized belief-propagation and circuit-level variants produce different values; the family therefore has no configuration-independent threshold \cite{Roffe2020BP}. Reinforcement-learning decoders have been demonstrated under several simplified phenomenological noise models, but the cited study does not define one transferable threshold across code geometry, distance, syndrome schedule, and physical noise \cite{Sweke2021RLDecoder}.

Learned decoders seek to absorb analog readout, leakage, space--time correlation, and device-specific nonstationarity. AlphaQubit used a recurrent-transformer architecture with $6.5\times10^6$ experimental Sycamore surface-code shots recorded across five device regions, two memory bases, and 13 round settings. A specific distance--basis--location task contains $650{,}000$ samples; the even/odd split leaves $325{,}000$ samples for fine-tuning and development and $325{,}000$ held out for final testing. Across the 13 round settings, one fine-tuning run uses $258{,}440$ unique training samples and $66{,}560$ development samples; repeated passes through that fixed set are not counted as new examples. Simulated pretraining used as many as $2.5\times10^9$ examples in the higher-distance scaling study. On experimental data the decoder obtained logical error per round $(2.901\pm0.023)\times10^{-2}$ at $d=3$ and $(2.748\pm0.015)\times10^{-2}$ at $d=5$, compared with $(3.028\pm0.023)\times10^{-2}$ and $(2.915\pm0.016)\times10^{-2}$ for the tensor-network baseline; the quoted uncertainties are the source-reported statistical intervals \cite{Bausch2024}. The source reports an unoptimized computation-time interval of approximately $10$--$100\,\mu$s per decoding round relative to the $1\,\mu$s target, without a mean or median estimate; this timing excludes final-answer latency after receipt of the last syndrome round. Table~\ref{tab:ai_decoders} is the quantitative entry for this study. A graph-neural-network decoder has shown approximately linear inference scaling with code space--time volume, performance above matching for the stated simulated circuit-level surface-code data, and parity with matching on experimental repetition-code data \cite{Lange2025GNN}. A recurrent neural decoder has also been evaluated for near-term surface-code experiments \cite{Varbanov2025}. Reinforcement-learning decoders remain predominantly simulation studies and must not be conflated with real-time fault-tolerant operation \cite{Sweke2021RLDecoder}.

Decoder selection is a multiobjective problem involving logical error, latency, memory, power, model-update cost, and robustness to distribution shift. A decoder that performs best on its training distribution can be inferior after calibration drift, leakage activation, a burst error, or a change in circuit schedule. Hardware integration therefore requires fixed maximum latency, confidence calibration, a fallback decoder, online monitoring of residuals, and retraining rules that do not invalidate the logical-error estimate. Table~\ref{tab:ai_decoders} compares the principal decoder families and shows that threshold, logical-error, and latency claims are commensurate only for a matched code, syndrome circuit, noise process, and implementation; learned accuracy gains must be evaluated together with deterministic inference latency and distribution-shift risk.

\begin{table}[tbp]
\caption{Decoder families for quantum error correction. Maximum-likelihood decoding (MLD), threshold, and logical-error comparisons apply only to the stated code, syndrome circuit, and noise assumptions.}
\label{tab:ai_decoders}
\centering
\begin{tblr}{colspec={Q[l,wd={0.15\linewidth}] Q[l,wd={0.19\linewidth}] Q[l,wd={0.23\linewidth}] Q[l,wd={0.20\linewidth}] Q[l,wd={0.22\linewidth}]}, cells={valign=t}, rowsep=1.35pt, colsep=1.0pt}
\toprule
Decoder family & Representative code and evidence & Accuracy or threshold statement & Computational property & Principal limitation \\
\midrule
MWPM \cite{Stephens2014} & Surface code; circuit-level numerical studies & Threshold $0.502(1)$--$1.140(1)\%$ across stated circuits and noise models & Polynomial matching; mature implementations & Requires a calibrated detector graph; correlations and hyperedges need approximation \\
Union--find \cite{Delfosse2021} & Two-dimensional toric code; numerical study & $9.9\%$ perfect-syndrome and $2.6\%$ faulty-syndrome thresholds & Worst-case $O[n\alpha(n)]$ & Thresholds are not circuit-level surface-code values; weighting and correlations require extensions \\
Tensor-network maximum-likelihood decoding (MLD) \cite{Bravyi2014Decoder} & Surface code; noiseless syndrome in the exact construction & Lower logical error than MWPM for reported tests with $\chi\ge4$ & Exact $O(n^2)$ for restricted noise; approximate $O(n\chi^3)$ otherwise & Bond-dimension cost and circuit-level syndrome noise \\
Belief propagation and ordered-statistics postprocessing \cite{Roffe2020BP} & Toric and sparse quantum LDPC codes; numerical studies & Representative belief-propagation plus ordered-statistics decoding (BP+OSD) threshold $9.9\pm0.2\%$ for the toric code with perfect syndrome; no configuration-independent family threshold & Iterative local messages; parallelizable & Short cycles, degeneracy, correlated faults, and convergence failures \\
Recurrent or transformer network \cite{Bausch2024,Varbanov2025} & Experimental Sycamore $d=3,5$ data; Pauli+ simulations through $d=11$ & AlphaQubit logical error per round $(2.901\pm0.023)\times10^{-2}$ ($d=3$) and $(2.748\pm0.015)\times10^{-2}$ ($d=5$), versus tensor-network $(3.028\pm0.023)\times10^{-2}$ and $(2.915\pm0.016)\times10^{-2}$ & Recurrent transformer; $6.5\times10^6$ shots recorded in aggregate; per distance--basis--location task: $650{,}000$ samples total, $325{,}000$ in the fine-tuning/development fold, and $258{,}440$ unique training samples; up to $2.5\times10^9$ simulated pretraining examples & Source-reported computation-time interval $10$--$100\,\mu$s per round, with no central estimate, versus a $1\,\mu$s target; final-answer latency excluded; training-distribution and circuit-transfer limitations \\
Graph neural network \cite{Lange2025GNN} & Simulated circuit-level surface code; experimental repetition-code data & Above matching in the stated simulations; on par with matching for the experimental repetition code & Inference approximately linear in code space--time volume & Training cost, experimental-data volume, and generalization to new circuits \\
Reinforcement learning \cite{Sweke2021RLDecoder} & Surface and related topological codes; predominantly simulation & No single-value threshold; configuration-dependent---see text & Sequential decision process & Exploration and training cost; no general real-time hardware superiority demonstrated \\
\bottomrule
\end{tblr}
\end{table}

\section{Quantum simulation}
\label{sec:simulation}

\subsection{Platforms, observables, and validation requirements}

Digital simulation represents a target Hamiltonian as $H=\sum_j c_jP_j$ and approximates or transforms $\exp(-\ii Ht)$; analog and annealing platforms implement an effective Hamiltonian directly \cite{Feynman1982,Lloyd1996,Georgescu2014}. The accepted scientific output is not the number of controllable sites but a declared observable---for example a structure factor, correlation length, spectrum, transport coefficient, phase boundary, or dynamical response---with a specified Hamiltonian, geometry, state preparation, readout model, sample count, uncertainty, finite-size treatment, and validation domain.

Neutral-atom and Rydberg arrays provide programmable geometry, strong interactions, and site-resolved readout. The 51-atom Ising experiment established transparent small-system comparisons \cite{Bernien2017}; later studies examined Kibble--Zurek scaling, two-dimensional phases, and quantum coarsening on arrays up to $16\times16$ in the cited configurations \cite{Keesling2019,Scholl2021,Ebadi2021,Manovitz2025}. A 104-atom digital experiment, comprising 72 data and 32 ancilla qubits, used mid-circuit measurement, feed-forward, tunable Floquet circuits, and error detection to probe the Kitaev honeycomb phase diagram and fermionic dynamics \cite{Evered2025Kitaev}. Its observables, postselection, and numerical noise model are specific to that circuit and cannot be merged with analog-array results. Trapped-ion simulators provide long-range spin couplings and high-fidelity site-resolved measurement; a 53-ion experiment observed nonequilibrium Ising dynamics and a dynamical phase transition \cite{Zhang2017IonSimulator}. Ultracold gases access Hubbard and continuum models: an approximately 80-site fermionic system observed antiferromagnetic correlations at $T/t\simeq0.25$ \cite{Mazurenko2017}, while a later cryogenic $^6$Li simulator prepared a strongly correlated region spanning about 340 lattice sites and used exact and approximate numerical comparisons in different doping regimes \cite{Xu2025Hubbard}.

Superconducting circuits implement spin, bosonic, and lattice models with fast control. A dissipatively stabilized photonic Bose--Hubbard array prepared an incompressible Mott state; its Author Correction adds omitted references but does not change the quantitative results used here \cite{Ma2019Mott}. A ten-qubit generalized Aubry--Andr\'e--Harper experiment mapped localized and topological dynamics with declared $U(1)$ postselection \cite{Li2023GAAH}. A $4\times4$ hard-core Bose--Hubbard array extracted correlation length and entanglement proxies across the spectrum; the study also shows how sampling requirements grow for volume-law states \cite{Karamlou2024}. Integrated photonics has simulated molecular vibronic spectra in a 16-mode processor, with source-reported reconstructed fidelities above $92\%$ for selected formic-acid and thymine cases \cite{Zhu2024Vibronic}. Ultracold polar molecules provide long-lived interacting pseudospins: approximately 2400 $^{87}$Rb$^{133}$Cs molecules showed $T_2^*=0.78(4)\,$s and a spin-echo lower bound $T_2>1.4\,$s at $95\%$ confidence in the reported magic trap \cite{Gregory2024Molecules}. This is enabling evidence for molecular simulation rather than a complete many-body application output.
%; the source provides public data under a Digital Object Identifier (DOI). 
Quantum annealers and coherent Ising architectures target optimization, sampling, and spin dynamics; interpretation depends on embedding, analog precision, schedule optimization, and the chosen classical solver \cite{King2022Annealing}.

\subsection{Layered validation and reproducibility}

Validation is necessarily layered. Hamiltonian and readout calibration establish the model implemented by the apparatus. Exact solutions, symmetries, conservation laws, or small-system calculations test tractable limits. Tensor-network, density-matrix renormalization group (DMRG), Monte Carlo, neural quantum states, or other controlled numerical methods extend comparison within their convergence domain. Multiple observables and parameter sweeps reduce the risk that one fitted quantity masks model error. Cross-platform or independent experimental replication is stronger still but remains uncommon.

Neural quantum states (NQS) are variational classical representations of amplitudes or density operators and must be included among the strongest documented baselines when their architecture matches the target state. For a wavefunction $\psi_{\boldsymbol\theta}(\boldsymbol\sigma)$,
\begin{equation}
E(\boldsymbol\theta)=
\frac{\langle\psi_{\boldsymbol\theta}|H|\psi_{\boldsymbol\theta}\rangle}
     {\langle\psi_{\boldsymbol\theta}|\psi_{\boldsymbol\theta}\rangle}
=\mathbb E_{p_{\boldsymbol\theta}(\boldsymbol\sigma)}
\!\left[E_{\rm loc}(\boldsymbol\sigma)\right],
\quad
p_{\boldsymbol\theta}(\boldsymbol\sigma)=
\frac{|\psi_{\boldsymbol\theta}(\boldsymbol\sigma)|^2}
     {\langle\psi_{\boldsymbol\theta}|\psi_{\boldsymbol\theta}\rangle},
\label{eq:nqs_variational}
\end{equation}
where $E_{\rm loc}(\boldsymbol\sigma)=\sum_{\boldsymbol\sigma'}H_{\boldsymbol\sigma\boldsymbol\sigma'}\psi_{\boldsymbol\theta}(\boldsymbol\sigma')/\psi_{\boldsymbol\theta}(\boldsymbol\sigma)$. Restricted Boltzmann machines, autoregressive convolutional or recurrent networks, and symmetry-equivariant networks trade representational flexibility against sampling and optimization cost \cite{Carleo2017NQS,Sharir2020NQS,HibatAllah2020NQS,Roth2023NQS}. Autoregressive models generate independent samples with explicit likelihoods, whereas nonautoregressive ansatzes commonly use Markov-chain Monte Carlo and can incur autocorrelation. The cost is set by sample count, network evaluation, stochastic-reconfiguration or gradient solves, and repeated optimization. Nontrivial sign or phase structure, multimodal distributions, ill-conditioned quantum Fisher matrices, and local minima can prevent convergence; low variational energy alone does not certify all observables \cite{Szabo2020NQSSign}. NQS are especially competitive for frustrated, two-dimensional, or volume-law regimes in which matrix-product states or sign-problem-free Monte Carlo are restricted, but they should be compared with DMRG and tensor networks, quantum Monte Carlo, sparse-state or sparse-operator methods, and exact diagonalization at the same Hamiltonian, size, symmetry sector, and accuracy target.

The 2025 Rydberg coarsening study illustrates this hierarchy: local-domain dynamics were compared with time-dependent variational principle (TDVP) calculations on $15\times15$ and $16\times15$ lattices; amplitude-mode dynamics and gaps were compared with DMRG and smaller-lattice calculations; energy conservation and multiple observables supplied internal checks \cite{Manovitz2025}. These tests support the declared observables and parameter regions, not a general assertion that all outputs are classically intractable. Table~\ref{tab:simulation} compares the accepted observables, operating conditions, validation routes, and principal inference limits for selected platforms.

\begin{table}[tbp]
\caption{Quantum-simulation operating points and validation limits organized by accepted observable; row order is nonranking and unlike Hamiltonians are not normalized. Numerical entries are source-transcribed unless explicitly identified as analytical. Time-dependent variational principle (TDVP) and density-matrix renormalization group (DMRG) are defined here because the table may float before their running-text discussion.}
\label{tab:simulation}
\centering
\begin{tblr}{colspec={Q[l,wd={0.15\linewidth}] Q[l,wd={0.195\linewidth}] Q[l,wd={0.20\linewidth}] Q[l,wd={0.23\linewidth}] Q[l,wd={0.21\linewidth}]}, cells={valign=t}, rowsep=1.3pt, colsep=1.0pt}
\toprule
Platform / configuration & Accepted observable or result & Scale and operating condition & Validation route & Principal inference limit \\
\midrule
Neutral-atom Rydberg, analog \cite{Manovitz2025} & Correlation growth, domain dynamics, structure factor, amplitude-mode frequency and damping & $16\times16$ array across a $(2+1)$-dimensional Ising transition & TDVP/DMRG, convergence and conservation checks, multiple observables & Numerical validation becomes restricted as size and entanglement grow; independent experimental replication absent \\
Neutral-atom Rydberg, digital \cite{Evered2025Kitaev} & Topological-state observables, Chern-number inference, fermionic dynamics, and interacting Fermi--Hubbard response & 104 atoms: 72 data and 32 ancillas; mid-circuit readout, feed-forward, Floquet gates, loss-based error detection & Exact free-fermion structure, symmetry constraints, measured strings, noise-model comparisons, and 68\% confidence intervals where reported & Postselection and model-assisted inference are circuit specific; no independent apparatus-level replication \\
Trapped ions \cite{Zhang2017IonSimulator} & Nonequilibrium Ising dynamics and dynamical phase transition & 53-ion chain with tunable long-range interactions & Controlled comparisons at smaller size and tractable limits; site-resolved readout & Long-range coupling calibration, finite size, decoherence, and one apparatus \\
Ultracold fermions \cite{Mazurenko2017,Xu2025Hubbard} & Antiferromagnetic correlations and doped Hubbard observables & Approximately 80-site and approximately 340-site strongly correlated regions in different experiments & Numerical methods selected by temperature, filling, and observable; cross-checks to exact regimes & Temperature extraction and conclusions away from half filling depend on model and numerical approximation \\
Superconducting circuits \cite{Ma2019Mott,Li2023GAAH,Karamlou2024} & Stabilized Mott state, localization/topology, correlation length and entanglement proxies & Bosonic array, ten-qubit chain, and $4\times4$ lattice in separate devices & Calibration, symmetry checks, exact/numerical comparison, declared postselection, sample studies & Different models and devices cannot be combined; sampling and tomography scale steeply \\
Integrated photonics \cite{Zhu2024Vibronic} & Molecular vibronic spectra; reconstructed fidelity above $92\%$ in selected cases & 16 optical modes for formic acid and thymine & Calibration and comparison with reconstructed target spectra & Loss, source statistics, reconstruction assumptions, and selected-mode fidelity \\
Ultracold molecules \cite{Gregory2024Molecules} & Long-lived rotational pseudospins and controlled dipolar decoherence & Approximately 2400 RbCs molecules in a magic optical trap & Ramsey/echo analysis, interaction tuning, many-body average-cluster-expansion comparison
%, public data with a Digital Object Identifier (DOI) 
& Primarily enabling interaction/coherence evidence rather than a completed large-scale simulation output \\
\bottomrule
\end{tblr}
\end{table}

AI-assisted Hamiltonian learning and experiment design are relevant because model discrepancy can dominate once direct classical verification becomes unavailable. For parameters $\boldsymbol\lambda$ and data $\mathcal D$,
\begin{equation}
p(\boldsymbol\lambda\mid\mathcal D)\propto p(\mathcal D\mid\boldsymbol\lambda)p(\boldsymbol\lambda),
\label{eq:bayes_simulation}
\end{equation}
but a posterior within one assumed model is not validation of the model class. Learned models must be tested on observables and parameter regions that were not used for calibration, and their uncertainty must be propagated to the scientific conclusion.

\section{Quantum communication and networking}
\label{sec:communication}
\label{sec:qkd}

Quantum communication uses nonorthogonal states, single-photon interference, entanglement, and quantum memories to distribute secret keys, states, or nonlocal correlations. Loss, detector background, phase stability, finite-size statistics, trust assumptions, and memory-interface efficiency determine the useful rate. Point-to-point QKD, trusted-node networks, entanglement distribution, and repeater architectures are therefore evaluated separately.
\subsection{Direct optical channels and key distribution}

For fiber attenuation $\alpha$ in dB km$^{-1}$, the transmissivity is
\begin{equation}
\eta_{\rm ch}(L)=10^{-\alpha L/10}.
\label{eq:fiber}
\end{equation}
At $1550\,$nm, the attenuation coefficient depends on fiber class and installation. International Telecommunication Union Telecommunication Standardization Sector (ITU--T) Recommendation G.654 reports achieved values of $0.15$--$0.19\,$dB km$^{-1}$ for cut-off-shifted ultralow-loss fiber \cite{ITUG6542024}; conventional installed links can be higher. The long twin-field links discussed below used ultralow-loss pure-silica-core fiber and therefore should not be interpreted using a generic standard-fiber interval. The two-way-assisted repeaterless secret-key capacity per optical mode, commonly called the Pirandola--Laurenza--Ottaviani--Banchi (PLOB) bound, is
\begin{equation}
K_{\leftrightarrow}(\eta)=-\log_2(1-\eta)
\simeq\frac{\eta}{\ln2},\qquad \eta\ll1,
\label{eq:plob}
\end{equation}
which provides the direct-channel reference \cite{Pirandola2017,Pirandola2020}.  For an asymptotic decoy-state Bennett--Brassard 1984 (BB84) implementation, a representative secret-key rate per emitted pulse is
\begin{equation}
r_K\geq q\Big[Q_1\bigl(1-h_2(e_1)\bigr)
-f_{\rm EC}Q_\mu h_2(E_\mu)\Big],
\label{eq:decoy_rate}
\end{equation}
where $q$ is the basis-sifting factor, $Q_1$ and $e_1$ are the single-photon gain and phase-error estimate, $Q_\mu$ and $E_\mu$ are the observed signal-state gain and quantum bit error rate (QBER), $f_{\rm EC}\geq1$ is the error-correction inefficiency, and $h_2$ is binary entropy \cite{Gisin2002,Scarani2009,Xu2020}.  Finite-size composable security subtracts terms determined by block length and failure parameters.  An achieved service rate also depends on source clock rate, detector background, authentication, classical post-processing, duty cycle, and availability.

\begin{figure}[tbp]
\centering
\includegraphics[alt={Three panels showing pure-loss secret-key capacity, direct finite-key QKD normalized by the nominal source clock with time bases identified per point, and twin-field QKD per sending-out pulse pair.},width=0.96\linewidth]{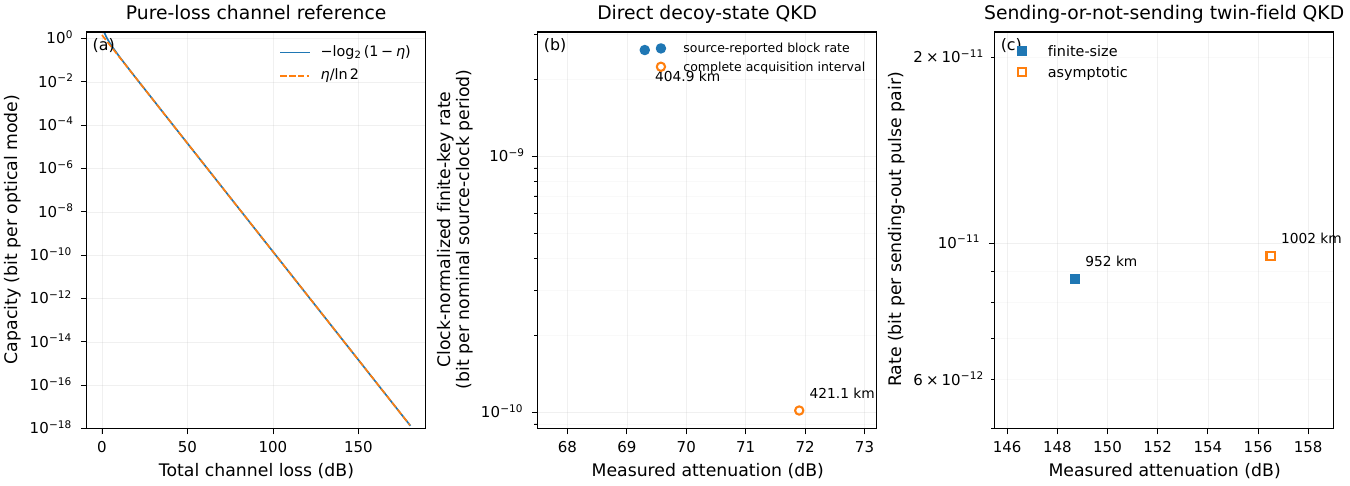}
\caption{Quantum-key-distribution quantities with explicit denominators. (a) Exact two-way pure-loss secret-key capacity per optical mode. (b) Direct finite-key decoy-state results divided by the nominal $2.5\,$GHz source clock. The filled $404.9\,$km marker uses the source-reported $6.5\,$bit s$^{-1}$ final rate associated with the reported $6.67\,$h block time; no separate transmission-active interval is given for that point. The open $421.1\,$km marker uses the complete-acquisition rate $22\,124/(24.2\,\mathrm{h})=0.2540\,$bit s$^{-1}$, or $1.016\times10^{-10}$ bit per nominal clock period; the distinct $0.484\,$bit s$^{-1}$ transmission-active rate is not plotted. (c) Sending-or-not-sending twin-field results retained as secret bit per sending-out pulse pair; filled and open squares distinguish finite-size and asymptotic analyses. The panels are not directly interchangeable because optical-mode, clock-normalized, and pulse-pair denominators, finite-size treatment, phase-reference overhead, and trust assumptions differ.}
\label{fig:qkd}
\end{figure}

BB84 uses nonorthogonal states; entanglement-based protocols use correlated measurements \cite{Bennett1984,Ekert1991}.  The denominator used for a rate comparison must be stated explicitly.  Equation~\eqref{eq:decoy_rate} is a protocol rate per emitted pulse.  By contrast, the long-distance experiment reports a nominal source clock $f_{\rm clk}=2.5\,$GHz and final secure throughput $R_{\rm sec}$ in bit s$^{-1}$.  Dividing by the nominal clock defines
\begin{equation}
r_{\rm clk}=\frac{R_{\rm sec}}{f_{\rm clk}},
\label{eq:qkd_pulse_norm}
\end{equation}
a clock-normalized finite-key rate that inherits the time denominator used for $R_{\rm sec}$; it is not automatically an elapsed-time or transmission-active yield. At $404.9\,$km the measured attenuation was $69.3\,$dB, and the source table reports a $6.67\,$h block time and final finite-key throughput $6.5\,$bit s$^{-1}$, giving $r_{\rm clk}=2.6\times10^{-9}$ bit per nominal source-clock period \cite{Boaron2018}. The source does not report a separate transmission-active interval for this operating point, so the value is retained as a source-reported block-rate normalization. At $421.1\,$km the measured attenuation was $71.9\,$dB and the source separately reports the complete acquisition interval and the transmission-active interval. The experiment used a three-state time-bin protocol, one decoy state, $\epsilon_{\rm sec}=\epsilon_{\rm cor}=10^{-9}$, superconducting detectors with source-reported $40$--$60\%$ efficiency and $0.1\,$Hz dark-count rate, and blockwise interruption for stabilization at the longest distance. Evaluating the published count and intervals, $22\,124/(24.2\,\mathrm{h})=0.2540\,$bit s$^{-1}$ reproduces and refines the source's approximate complete-acquisition rate of $0.25\,$bit s$^{-1}$, whereas $22\,124/(12.7\,\mathrm{h})=0.484\,$bit s$^{-1}$ is the distinct transmission-active rate \cite{Boaron2018}. Figure~\ref{fig:qkd}(b) therefore shows two rates with the same nominal-clock normalization but different documented time bases: the source-reported $404.9\,$km block rate and the reconstructed $421.1\,$km complete-acquisition rate. The transmission-active $421.1\,$km value is not plotted. Their ratio, $1.91$, is a denominator effect already implicit in the source report, not a correction to it; all arithmetic inputs are stated here. Treating the printed $24.2\,$h and $12.7\,$h durations as rounded to the nearest $0.1\,$h gives deterministic rounding intervals $0.2534$--$0.2545\,$bit s$^{-1}$ and $0.4820$--$0.4858\,$bit s$^{-1}$, respectively. These ranges are not statistical confidence intervals, and the source does not provide a covariance model for formal propagation.

Twin-field QKD uses single-photon interference at an untrusted middle station and has a different channel-use convention \cite{Lucamarini2018}.  The 1000-km sending-or-not-sending experiment defined its rate per \emph{sending-out pulse pair}; that convention is retained rather than converted to an optical mode or single emitted pulse.  The measured total attenuation was $148.7\,$dB at $952\,$km and $156.5\,$dB at $1002\,$km, corresponding to path-averaged coefficients $148.7/952=0.15620\,$dB km$^{-1}$ and $156.5/1002=0.15619\,$dB km$^{-1}$. These values characterize the ultralow-loss experimental links and lie below the range commonly quoted for standard telecom installations. The source reported $8.75\times10^{-12}$ bit per pulse pair for the finite-size $952\,$km result and $9.53\times10^{-12}$ bit per pulse pair for the asymptotic $1002\,$km result \cite{Liu2023TFQKD}.  The physical modulation rate was $1\,$GHz, the effective signal rate in the long-distance configuration was $351\,$MHz, the phase estimate was refreshed every $500\,$ms, and the finite-size calculation used the complete test length with the source-reported composable security parameters.  Because pulse, pulse-pair, and optical-mode normalizations are not interchangeable, Fig.~\ref{fig:qkd} displays them in separate panels.

Satellite QKD and entanglement distribution avoid exponential terrestrial-fiber attenuation over most of the path.  In the dual-downlink entanglement experiment, ground stations separated by $1203\,$km received coincidences at an average $1.1\,$Hz; the reconstructed two-photon state fidelity was $0.869\pm0.085$, and the Bell parameter was $2.37\pm0.09$ under the source's analysis conditions \cite{Yin2017}.  A separate decoy-state satellite-to-ground QKD experiment reported secret-key rates at the kilobit-per-second scale over slant ranges up to approximately $1200\,$km \cite{Liao2017}; a later trusted-node space-to-ground network integrated satellite and terrestrial links \cite{Chen2021QKD}.  These rates apply during finite satellite passes rather than continuous service. The system budget is controlled by telescope aperture, pointing and tracking, atmospheric transmission, background, detector timing, pass duration, weather, and trusted ground infrastructure.

Integrated photonics addresses replication and optical referencing.  A 2026 proof-of-principle network implemented a four-intensity sending-or-not-sending twin-field quantum key distribution protocol in a measurement-device-independent star topology: 20 InP transmitters formed ten wavelength-paired links, each using two approximately $185\,$km upstream fiber spools to an untrusted central interference node referenced by a Si$_3$N$_4$ microcomb \cite{Zheng2026QKD}.  For $N_{\rm tot}=10^{12}$ sent protocol bins per pair, the source reported composable finite-key rates $R$ in secret bits per pulse.  The attenuation coefficient used to draw the source's direct-link PLOB reference is not tabulated in its numerical-parameter table.  Reconstruction of the dashed reference and the reported $51.5\%$--$251.4\%$ enhancements gives the nominal convention
\begin{equation}
\alpha_{\rm PLOB}=0.200\ \mathrm{dB\,km^{-1}},\quad
A_{\rm PLOB}=\alpha_{\rm PLOB}(370\ \mathrm{km})=74.0\ \mathrm{dB},\quad
K_{\leftrightarrow}=5.743\times10^{-8}\ \mathrm{bit\,mode^{-1}},
\label{eq:zheng_plob_reconstruction}
\end{equation}
with $\eta_{AB}=10^{-A_{\rm PLOB}/10}=3.981\times10^{-8}$.  The plotted comparison treats one sent protocol bin as the direct-link channel use corresponding to the rate denominator.  On that convention, the stated enhancement range corresponds to $R=8.70\times10^{-8}$--$2.02\times10^{-7}\ \mathrm{bit\,pulse^{-1}}$, and all ten links lie above the nominal point-to-point repeaterless reference.  The accessible article and supplement identify spooled telecom-C-band fiber near $1550\,$nm but do not specify a commercial fiber type.  The source's channel-characterization table separately reports fiber-only losses of $34.49\pm0.02\,$dB over $185.131\,$km and $35.11\pm0.03\,$dB over $185.299\,$km, giving $69.60\pm0.04\,$dB over $370.430\,$km, an effective path-averaged coefficient $0.18789\pm0.00010\ \mathrm{dB\,km^{-1}}$, and $K_{\leftrightarrow}=1.582\times10^{-7}\ \mathrm{bit\,mode^{-1}}$.  Under this measured-fiber convention, the reported finite-key range straddles rather than uniformly exceeds the direct-link capacity.  Filter and circulator insertion losses were characterized separately and are not folded into either fiber-only coefficient.  Thus the ``all ten above PLOB'' statement is specific to the source's nominal $0.200\ \mathrm{dB\,km^{-1}}$ equivalent-channel convention; over $370\,$km, changing $\alpha$ by $0.01\ \mathrm{dB\,km^{-1}}$ changes the reference capacity by a factor $10^{0.37}=2.34$.  The comparison does not imply violation of the applicable capacity bound for the two-arm network.  Pairwise runs were sequential, demonstrating integrated multi-client hardware and shared phase referencing rather than simultaneous ten-link service.

\subsection{Memories, repeaters, and distributed quantum functions}

A repeater node combines a stationary qubit or ensemble, optical interface, memory, Bell-state measurement, feed-forward, synchronization, and often wavelength conversion.  For a symmetric two-photon elementary link,
\begin{equation}
p_{\rm succ}=p_{s,A}p_{s,B}p_B
\prod_{j=A,B}\eta_{c,j}\eta_{{\rm conv},j}
\eta_{{\rm ch},j}\eta_{{\rm det},j},
\label{eq:link}
\end{equation}
where $p_s$ is source probability, $p_B$ Bell-measurement success, and the $\eta$ terms represent collection, conversion, channel, and detector efficiencies.  The raw rate is $R_E=f_{\rm rep}p_{\rm succ}$.  Waiting-time statistics, memory decoherence, multiplexing, swapping, purification or error correction, routing, and node availability further reduce end-to-end performance.
For a geometric sequence of independent attempts, the mean elementary-link waiting time is approximately
\begin{equation}
\langle t_{\rm wait}\rangle\simeq\frac{1}{f_{\rm rep}p_{\rm succ}},
\label{eq:waiting_time}
\end{equation}
so useful swapping requires memory coherence and storage fidelity over a distribution whose mean and tail both exceed the communication latency.  In a multi-link repeater the maximum of several stochastic waiting times, not a single-link mean, sets the memory requirement.

A metropolitan-scale multiplexed repeater experiment reported remote-memory entanglement over $14.5\,$km with Bell-state fidelity $78.6\%\pm2.0\%$ and Bell nonlocality \cite{Zhu2026Repeater}; the source-reported interval is retained with coverage \NR{} where not stated in the accessible text.  Entanglement-assisted nonlocal optical interferometry over a $1.55\,$km baseline reported Bell-state fidelity $F=0.63(3)$ and an average data-collection rate of about $12\,$mHz, set by a $0.41\,$Hz entanglement rate and additional erasure and heralding probabilities \cite{Stas2026}.  The source identifies the relevant fidelity and visibility error bars as one standard deviation.  These experiments establish memory-assisted and application-specific network primitives.  They do not yet provide a routed repeater service whose rate and availability exceed direct transmission.

The central scaling problem is simultaneous optimization: memory lifetime must exceed stochastic entanglement waiting time; write/read efficiency and wavelength conversion must remain high; multiphoton and background errors must remain below the fidelity budget; and multiplexing must increase rate faster than it increases switch loss, control complexity, and memory demand.  A repeater milestone is therefore an end-to-end entanglement rate and fidelity exceeding a direct or trusted-node baseline under the same channel, security, clock-rate, and availability assumptions.

\subsection{Microwave links and quantum transduction}

Superconducting circuits and many spin systems operate at microwave frequencies, where thermal occupation prohibits room-temperature propagation without overwhelming added noise.  Microwave-to-optical transduction seeks to connect cryogenic processors to low-loss optical channels.  The relevant metrics are bidirectional conversion efficiency, added noise referred to the input, instantaneous bandwidth, pump-induced heating, isolation, and compatibility with qubit coherence.  High efficiency without low added noise is insufficient, and a narrow high-cooperativity demonstration is not an interconnect unless it preserves a nonclassical state at a useful rate.

\subsection{Field networks and operational evidence}

Distance records and field operation answer different questions. A provider-authored report describes a 46-node metropolitan QKD network with 40 user nodes, three trusted relays, and three optical switches, operated for 31 months \cite{Wang2021MAN}. Eleven user nodes were included in the long-duration robustness record; key rates were sampled every $30\,$s and averaged by month, with source-reported link ranges of approximately $6$--$60.5\,$kbit s$^{-1}$. The network supported one-time-pad voice, text, and file transfer and up to 11 simultaneous user pairs in the reported demonstration. This is substantial site-specific field evidence, but it is not an independent availability assessment or a universal service-level commitment: the paper does not publish a common uptime denominator, maintenance distribution, incident record, or externally verified endpoint-assurance boundary.

An integrated photonic secure-communication system combined a chip-based transmitter, real-time QKD processing, a $4\,$Gbit s$^{-1}$ on-chip quantum random-number generator (QRNG), and line-rate encryption on metropolitan fiber \cite{Paraiso2021}. These results demonstrate system integration for named configurations. A network claim additionally requires the provider and user boundary, key-consumption workload, authentication method, availability, planned maintenance, incident response, and endpoint security.

Implementation assurance is distinct from optical-protocol performance. European Telecommunications Standards Institute (ETSI) Group Specification (GS) QKD 016 specifies a Common Criteria protection profile for a pair of prepare-and-measure QKD modules through their final secret-key output \cite{ETSIQKD016}. Conventional migration also includes the finalized NIST post-quantum standards for module-lattice key encapsulation and module-lattice and stateless-hash signatures \cite{NISTFIPS203,NISTFIPS204,NISTFIPS205}. These standards do not validate a particular QKD link; they define assurance and interoperability contexts against which deployed cryptographic systems must be evaluated.

Table~\ref{tab:communication} organizes the selected links by their native output and denominator. Its main conclusion is that distance alone does not order direct QKD, twin-field QKD, memory-assisted entanglement, satellite links, or field networks: loss, finite-size and trust model, memory-interface efficiency, and availability constrain different functions.

\begin{table}[tbp]
\caption{Selected communication and networking configurations. Rates retain the source denominator and security or fidelity convention. Each row refers to one apparatus and operating condition.}
\label{tab:communication}
\centering
\begin{tblr}{colspec={Q[l,wd={0.16\linewidth}] Q[l,wd={0.24\linewidth}] Q[l,wd={0.20\linewidth}] Q[l,wd={0.17\linewidth}] Q[l,wd={0.20\linewidth}]}, cells={valign=t}, rowsep=1.3pt, colsep=1.3pt}
\toprule
Configuration & Measurand and result & Security or uncertainty convention & Evidence scope & Principal limiting mechanism or missing evidence \\
\midrule
Direct decoy-state QKD \cite{Boaron2018} & $6.5\,$bit s$^{-1}$ at $69.3\,$dB; at $71.9\,$dB, source $\approx0.25\,$bit s$^{-1}$, elapsed-time evaluation $0.2540\,$bit s$^{-1}$, and transmission-active evaluation $0.484\,$bit s$^{-1}$ & Finite-key composable security with source-stated failure parameters & Laboratory fiber spools; full-block throughput includes stated stabilization and processing & Deployed-fiber campaign, endpoint assurance, users, availability, maintenance, and failures \\
Sending-or-not-sending twin-field QKD \cite{Liu2023TFQKD} & $8.75\times10^{-12}$ bit per sending-out pulse pair at $148.7\,$dB ($0.15620\,$dB km$^{-1}$ path average); $9.53\times10^{-12}$ at $156.5\,$dB ($0.15619\,$dB km$^{-1}$) & First value finite size; second asymptotic & Laboratory long-fiber configuration with phase reference & Maintained finite-block service at the largest loss, endpoint evidence, and phase-reference availability \\
Satellite entanglement and QKD \cite{Yin2017,Liao2017} & Dual-downlink entanglement over $1203\,$km: $1.1\,$Hz average coincidence rate, state fidelity $0.869\pm0.085$, Bell parameter $2.37\pm0.09$; separate decoy-state QKD at kilobit-per-second scale over slant ranges up to $\sim1200\,$km & Coincidence rate per effective pass time; source-reported state and Bell intervals; QKD rate in secret bit s$^{-1}$ during a pass & Orbital field demonstrations during finite satellite passes & Pass and weather availability, endpoint trust, and sustained key delivery \\
Remote-memory networking \cite{Zhu2026Repeater,Stas2026} & $78.6\%\pm2.0\%$ Bell-state fidelity over $14.5\,$km; separate $0.63(3)$ fidelity at about $12\,$mHz & Source-reported intervals and protocol-specific heralding & Application-specific network primitives in separate experiments & Write/read and conversion loss, waiting-time tails, swapping, routing, and node availability \\
Hefei metropolitan QKD network \cite{Wang2021MAN} & Provider reports 31 months; 11 nodes in robustness record; approximately $6$--$60.5\,$kbit s$^{-1}$ link ranges & Provider-authored field record; no common independent availability denominator & Deployed 46-node trusted-node network with named applications & Independent service-level availability, maintenance and failure distributions, endpoint assurance \\
\bottomrule
\end{tblr}
\end{table}

\section{Quantum sensing and imaging}
\label{sec:sensing}

Quantum sensors encode a field, force, displacement, rotation, or material parameter into a phase, population, quadrature, or correlation. Their performance is determined by the complete transfer function rather than by a noise floor alone. For a linearized readout,
\begin{equation}
y(\omega)=H(\omega)x(\omega)+n(\omega),\qquad
S_{x,\mathrm{eq}}^{1/2}(\omega)=\frac{S_n^{1/2}(\omega)}{|H(\omega)|},
\label{eq:sensor_transfer}
\end{equation}
so sensitivity must be reported with bandwidth, dynamic range, calibration, drift, and the operating point. With $[X,P]=\ii/2$, the vacuum variance is $V_{\rm vac}=1/4$. For the unnormalized variance $V_X\equiv\operatorname{Var}(X)$, linear loss of total efficiency $\eta$ gives $V_{X,\rm obs}=\eta V_{X,\rm in}+(1-\eta)V_{\rm vac}$. Defining the lowercase normalized variance,
\begin{equation}
v_X\equiv\frac{V_X}{V_{\rm vac}},\qquad V_{\rm vac}=\frac14,
\qquad v_{X,\rm obs}=\eta v_{X,\rm in}+1-\eta,
\label{eq:squeezing_loss}
\end{equation}
vacuum corresponds to $v_X=1$, and $\eta$ is the total source-to-estimator efficiency. Quantum enhancement is therefore useful only when technical noise and loss remain below the quantum contribution for the relevant frequency band and measurement protocol.
\subsection{Atom interferometry and inertial sensing}

\subsubsection{Signal model and architecture}

A light-pulse Mach--Zehnder interferometer uses a $\pi/2-\pi-\pi/2$ sequence to split, redirect, and recombine atomic wave packets.  The leading acceleration phase is
\begin{equation}
\Delta\phi_a=\bm{k}_{\rm eff}\cdot\bm a T^2,
\label{eq:atom_interferometer_phase}
\end{equation}
where $\bm a$ is the acceleration of the atomic wave packet relative to the optical phase reference, usually the retroreflection mirror, in units of $\mathrm{m\,s^{-2}}$; $\bm{k}_{\rm eff}$ is the effective Raman or Bragg wave vector, and $T$ is the pulse separation \cite{Cronin2009,Bongs2019}.  Equivalently, with $a_{\parallel}\equiv\hat{\bm{k}}_{\rm eff}\!\cdot\!\bm a$, the phase is $\Delta\phi_a=k_{\rm eff}a_{\parallel}T^2$. Counterpropagating beams near $780\,$nm give $k_{\rm eff}=4\pi/\lambda\simeq1.61\times10^7\,$m$^{-1}$.  Thus $1\,\mu$Gal $=10^{-8}\,$m s$^{-2}$ gives $0.161\,$rad at $T=1\,$s.  Rotation and gravity-gradient phases scale approximately as $2\bm{k}_{\rm eff}\cdot(\bm\Omega\times\bm v)T^2$ and $k_{\rm eff}\Gamma L T^2$, respectively.

A deployable instrument includes a vacuum package, atomic source, cooling and state preparation, Raman or Bragg optics, a stable reference mirror, vibration sensing, magnetic control, population readout, and a joint phase estimator.  The reference mirror is part of the inertial measurement: mirror motion contributes phase indistinguishable from acceleration unless independently measured or rejected.

For a three-pulse sequence with instantaneous pulses, the acceleration transfer function may be written
\begin{equation}
H_a(\omega)=\frac{4 k_{\rm eff}}{\omega^2}
\sin^2\!\left(\frac{\omega T}{2}\right),
\label{eq:atom_interferometer_transfer}
\end{equation}
which approaches $k_{\rm eff}T^2$ at low frequency and rolls off above a scale of order $1/T$. If $\sigma_\phi$ is the single-shot ensemble phase root-mean-square uncertainty and samples are statistically independent with cycle time $T_c$, define the single-shot acceleration uncertainty $\sigma_a=\sigma_\phi/(k_{\rm eff}T^2)$ and the white-noise Allan-deviation coefficient
\begin{equation}
\mathcal{S}_a\equiv \sigma_a\sqrt{T_c}
=\frac{\sigma_\phi\sqrt{T_c}}{k_{\rm eff}T^2},
\qquad
\sigma_{A,a}(\tau)=\frac{\mathcal{S}_a}{\sqrt{\tau}}.
\label{eq:atom_interferometer_sensitivity}
\end{equation}
For the same discrete white sequence, the two-sided power spectral density is $S_a^{(2)}=\sigma_a^2T_c=\mathcal{S}_a^2$, whereas the one-sided convention for $f\geq0$ is $S_a^{(1)}=2\sigma_a^2T_c=2\mathcal{S}_a^2$ and therefore $[S_a^{(1)}]^{1/2}=\sqrt{2}\,\mathcal{S}_a$. The campaign values quoted below follow the source convention $\mathcal{S}_a$, obtained by extrapolating the white Allan deviation to $\tau=1\,$s; they are not one-sided amplitude spectral densities. Equations~\eqref{eq:atom_interferometer_transfer} and \eqref{eq:atom_interferometer_sensitivity} make explicit the sensitivity--bandwidth--dynamic-range trade: increasing $T$ improves the low-frequency scale factor while reducing update rate and increasing phase ambiguity and platform-motion coupling.

\subsubsection{Performance, classical comparison, and scaling limits}

The field study evaluated the named instrument family Absolute Quantum Gravimeter (AQG), specifically the first unit AQG-A01, and provides informative co-located evidence but not a fully matched instrument comparison \cite{Menoret2018}. The source uses AQG as the instrument-family name and AQG-A01 as the identifier of the tested unit, rather than as a generic abbreviation for every atom-interferometric gravimeter. The AQG measured continuously for one month and had a campaign white-noise Allan coefficient $\mathcal{S}_a$ of approximately $750\,$nm s$^{-2}$ Hz$^{-1/2}$, with a one-day standard deviation of $9.4\,$nm s$^{-2}$. The FG5 made two discrete absolute measurements; the Allan-deviation comparison used one $32\,$h record comprising $32$ sets of $100$ drops, with drops separated by $10\,$s and sets by $1\,$h. The authors plotted those drops as an effective $36\,$s sequence, which includes substantial dead time, and obtained $\mathcal{S}_a\simeq450\,$nm s$^{-2}$ Hz$^{-1/2}$.  They estimated that uninterrupted $10\,$s sampling would improve that coefficient by $\sqrt{3.6}\simeq1.9$ to about $240\,$nm s$^{-2}$ Hz$^{-1/2}$.  Co-location therefore supports consistency and drift checks, but the unequal duration, duty cycle, and sampling procedure do not support a like-for-like sensitivity ranking.  The system-level distinction is continuous field monitoring versus discrete absolute-reference measurements.  Compact grating magneto-optical-trap and photonic-integration approaches target higher data rate and reduced optical volume \cite{Lee2022}; a cold-atom gyroscope has also been operated in orbit, where vibration, platform rotation, thermal variation, and autonomous control dominate the system design \cite{Li2025}.

Sensitivity improves as $k_{\rm eff}T^2$, but longer $T$ increases cycle time, instrument length, phase ambiguity, and susceptibility to wave-front aberration, Coriolis terms, vibration, magnetic gradients, spontaneous emission, laser phase noise, atom loss, and detection nonlinearity.  Large-momentum-transfer pulses increase $k_{\rm eff}$ at the cost of optical power and pulse infidelity.  Navigation-grade systems require hybrid classical inertial sensors to provide bandwidth and dynamic range while the atom interferometer constrains long-term bias.  The relevant comparison is therefore mission trajectory error or gravity uncertainty under matched motion, not the atom shot-noise floor alone.

\subsection{Magnetic, spin, and radio-frequency sensing}

\subsubsection{Atomic and superconducting magnetometers}

Atomic magnetometers infer magnetic field from Zeeman precession, $\omega_L=\gamma B$.  For $N$ polarized spins, coherence time $T_2$, readout contrast $C$, and averaging time $\tau$, a representative projection-noise scale is
\begin{equation}
\delta B\sim\frac{1}{\gamma C\sqrt{N T_2\tau}}.
\label{eq:mag_sensitivity}
\end{equation}
The usable response additionally depends on optical-pumping rate, spin-exchange and spin-destruction rates, cell geometry, probe detuning, photon shot noise, and the transfer function of magnetic shielding and feedback.  Spin-exchange-relaxation-free (SERF) operation requires the Larmor rate to remain below the spin-exchange collision rate and consequently can provide a low magnetic-field noise floor only over a restricted field range.  A shielded SERF experiment reported $0.54\,$fT Hz$^{-1/2}$ sensitivity in a $0.3\,$cm$^3$ active volume \cite{Kominis2003}.  The relevant classical or competing references depend on the application: superconducting quantum interference devices (SQUIDs) for cryogenic low-field sensing, mature optically pumped magnetometers for biomagnetism, fluxgates for vector field operation, and induction coils for finite-frequency signals.  Field deployment is controlled by vector calibration, heading error, Johnson noise, sensor-to-source distance, motion rejection, and dynamic range rather than the shielded scalar noise floor alone.

\subsubsection{Color-center sensors}

The negatively charged nitrogen-vacancy (NV) center in diamond is a room-temperature spin-1 defect with zero-field splitting $D\simeq2.87\,$GHz.  Optical pumping initializes the spin, microwave fields drive the $m_s=0\leftrightarrow m_s=\pm1$ transitions, and spin-dependent fluorescence provides readout.  In continuous-wave optically detected magnetic resonance, a useful photon-shot-noise estimate is
\begin{equation}
\eta_B^{\rm CW}\equiv S_B^{1/2}\simeq
\frac{4}{3\sqrt{3}}\frac{\Delta f}{\gamma_e^{(f)}C\sqrt{R}}
=
\frac{4}{3\sqrt{3}}\frac{\Delta\omega}{\gamma_e^{(\omega)}C\sqrt{R}},
\label{eq:nv_sensitivity}
\end{equation}
where $\Delta f$ is the Lorentzian full width at half maximum in hertz, $\Delta\omega=2\pi\Delta f$ is the angular-frequency linewidth, $C$ is the optical contrast, and $R$ is the detected photon rate in s$^{-1}$. The frequency-unit and angular-frequency gyromagnetic ratios are $\gamma_e^{(f)}=g_e\mu_B/h\simeq28.0\,$GHz T$^{-1}$ and $\gamma_e^{(\omega)}=g_e\mu_B/\hbar=2\pi\gamma_e^{(f)}$, respectively. Eq.~\eqref{eq:nv_sensitivity} is the ideal continuous-wave, Lorentzian, photon-shot-noise expression evaluated at maximum slope; line-shape distortion and technical noise change its prefactor \cite{Doherty2013,Rondin2014,Barry2020}.  NV sensors generally do not exceed optimized SERF or SQUID instruments in bulk field units; their advantage is nanometer-scale proximity, vector reconstruction, and room-temperature spatial imaging.  Surface-spin noise, charge-state conversion, strain, spectral diffusion, microwave inhomogeneity, laser heating, photon collection, and calibration of the sensor--sample stand-off dominate the accepted magnetic image.

\subsubsection{Rydberg radio-frequency electrometry}

Rydberg-atom electrometry uses electromagnetically induced transparency (EIT) and Autler--Townes splitting.  For a radio-frequency (RF) field coupling two Rydberg states,
\begin{equation}
\Omega_{\rm RF}=\frac{d_{\rm RF}E}{\hbar},
\qquad E=\frac{\hbar\Omega_{\rm RF}}{d_{\rm RF}},
\label{eq:rydberg_field}
\end{equation}
where $\Omega_{\rm RF}$ is the on-resonance angular Rabi frequency in rad s$^{-1}$ and $d_{\rm RF}=\langle r_2|\bm d\!\cdot\!\hat{\bm e}_{\rm RF}|r_1\rangle$ is the dipole matrix element projected onto the RF polarization.  If an ordinary frequency splitting $\nu_{\rm RF}$ is measured instead, $\Omega_{\rm RF}=2\pi\nu_{\rm RF}$.  The conversion therefore depends on the quantization axis, polarization, transition selection rules, Zeeman or hyperfine sublevels, detuning convention, and the line-shape model used to infer the splitting \cite{Sedlacek2012,Holloway2014,Adams2020}. In a rubidium vapor cell using a bright resonance within an electromagnetically induced-transparency window, the source reported sensitivity of approximately $30\,\mu\mathrm{V\,cm^{-1}\,Hz^{-1/2}}$ and detection of fields as small as approximately $8\,\mu\mathrm{V\,cm^{-1}}$; no statistical interval was attached to these approximate headline values \cite{Sedlacek2012}. Spatial field nonuniformity and calibration of the optical probe and coupling response must be included in the field estimate.  Laser linewidth, Doppler and transit-time effects, power broadening, collisions, blackbody transitions, and RF multipath broaden or bias the transfer function.  A communications or spectrum-sensing architecture must be compared with an antenna--low-noise-amplifier--mixer chain using input-referred field or noise temperature, instantaneous bandwidth, linearity, dynamic range, polarization and angle-of-arrival accuracy, calibration interval, optical power, and survivability.  The function most directly supported by current evidence is traceable RF field metrology and field mapping; broadband receiver substitution is not yet demonstrated under matched resource constraints.

\subsection{Quantum imaging and detection}

Quantum optical imaging uses single-photon timing, photon-number statistics, squeezed quadratures, spatial or temporal correlations, or nonlinear wavelength conversion to estimate a scene parameter under a constrained photon budget \cite{Moreau2019}. For incident photon number $N_\gamma$ at the declared dose boundary and end-to-end detection efficiency $\eta$, the detected count is $N_{\gamma,\mathrm{det}}=\eta N_\gamma$; a shot-noise-limited coherent measurement therefore has signal-to-noise ratio proportional to $\sqrt{N_{\gamma,\mathrm{det}}}=\sqrt{\eta N_\gamma}$. A quantum advantage must therefore be reported at fixed incident or absorbed dose, optical bandwidth, field of view, acquisition time, and reconstruction prior. Loss before the informative interaction wastes dose; loss after the interaction reduces Fisher information and can erase the nonclassical gain.

Frequency-dependent squeezed readout is integrated into Advanced LIGO. The first full-scale implementations reported amplitude-noise reductions of $4.0\,$dB at Hanford and $5.8\,$dB at Livingston near $1\,$kHz, corresponding to factors $10^{4.0/20}=1.58$ and $10^{5.8/20}=1.95$; the source rounded these factors to $1.6$ and $1.9$, respectively \cite{Ganapathy2023}. The same study reported a $15$--$18\%$ binary-neutron-star range increase relative to no squeezing. Subsequent Livingston operation reduced quantum noise below the standard quantum limit by up to $3\,$dB between $35$ and $75\,$Hz \cite{Jia2024}. During the fourth observing run, the A+ filter-cavity architecture enabled broadband quantum-noise reduction up to $5.2\,$dB at Hanford and $6.1\,$dB at Livingston. The detector-performance paper reports site duty factors of $65.0\%$ and $71.2\%$, with $52.6\%$ dual coincidence, over O4a and the first six months of O4b \cite{Capote2025LIGOO4}.

The reconstruction performed for this review uses the official Gravitational Wave Open Science Center O4a boundaries from Global Positioning System (GPS) time $1368975618$ to $1389456018$ and the public Hanford (H1) and Livingston (L1) \texttt{DMT-ANALYSIS\_READY:1} segment lists \cite{GWOSCO4a2025}. After clipping all intervals to that boundary, merging overlaps, and intersecting the two site lists, H1 contains $812$ intervals totaling $13\,821\,685\,$s, L1 contains $559$ intervals totaling $14\,124\,552\,$s, and the dual-site intersection contains $944$ intervals totaling $10\,935\,959\,$s. For the audited duration $T=20\,480\,400\,$s,
\[
D_{\rm H1}=0.67487,\qquad D_{\rm L1}=0.68966,\qquad D_{\rm H1\cap L1}=\frac{10\,935\,959}{20\,480\,400}=0.53397.
\]
No additional event-selection, source-class, or squeezing-amplitude criterion is applied. These are fractions of O4a elapsed time carrying the public analysis-ready flag, not event-acceptance probabilities or the fraction of time at the maximum reported squeezing level. Segment endpoints are integer GPS seconds and are treated as half-open intervals $[a,b)$; the quoted fractions are exact for the declared lists and preprocessing rather than statistical estimates with confidence intervals. Alternative data-quality flags, calibration versions, or audit boundaries define different data products and are not propagated as random uncertainty. Relative to the broader source interval, the O4a-only H1 fraction moves upward while the L1 fraction moves downward because the time windows and public data-quality boundaries differ; none of the pairs is a competing estimate of one common interval. A $3\,$dB variance reduction corresponds to a factor of two in noise power or $\sqrt{2}$ in amplitude sensitivity only where classical displacement, thermal, seismic, and coating noise remain below the quantum contribution. The realized gain depends on source squeezing, injection and output losses, mode matching, filter-cavity phase, interferometer control, calibration, and the fraction of elapsed time for which calibrated analysis-ready strain is available.

Single-photon and correlated-photon imagers can provide depth resolution, background rejection, or wavelength conversion that is difficult to reproduce with direct detection.  A phase-microscopy experiment using spatially entangled photons reported $2.76\,\mu\mathrm{m}$ spatial resolution, phase sensitivity near $\lambda/100$, and $100\,$fW illumination \cite{Zhang2026Imaging}.  These are laboratory operating conditions rather than a general imaging envelope: field of view, dwell time, collection numerical aperture, sample scattering, detector dead time, and reconstruction bias determine whether the result exceeds a classical low-light microscope at matched dose.

Table~\ref{tab:sensors} connects each sensor result to its bandwidth, drift, calibration, and operating duty factor. It shows that squeezed interferometry is integrated into a maintained observatory, whereas gravimetry, magnetometry, microscopy, and radio-frequency sensing are limited by distinct transfer-function and field-operation terms rather than by one sensitivity number.

\begin{table}[tbp]
\caption{Selected sensing and imaging configurations. Values apply only to the cited apparatus and condition; sensitivity, bandwidth, drift, and duty factor are distinct quantities.}
\label{tab:sensors}
\centering
\begin{tblr}{colspec={Q[l,wd={0.17\linewidth}] Q[l,wd={0.23\linewidth}] Q[l,wd={0.22\linewidth}] Q[l,wd={0.18\linewidth}] Q[l,wd={0.18\linewidth}]}, cells={valign=t}, rowsep=1.5pt, colsep=1.3pt}
\toprule
Configuration & Measurand and result & Uncertainty interpretation & Evidence scope / condition & Principal limiting mechanism \\
\midrule
Absolute Quantum Gravimeter (AQG), unit AQG-A01 / FG5 \cite{Menoret2018} & AQG: white-noise Allan coefficient $\mathcal{S}_a\simeq750\,$nm s$^{-2}$ Hz$^{-1/2}$ and $9.4\,$nm s$^{-2}$ at one day; FG5: $\mathcal{S}_a\simeq450\,$nm s$^{-2}$ Hz$^{-1/2}$ as operated & Descriptive campaign statistics; coefficients follow the source Allan-deviation convention, not a one-sided amplitude spectral density; no common confidence interval & AQG continuous month; FG5 discrete measurements and one $32\,$h record & Vibration, wave fronts, bias, dynamic range, dead time, and maintenance; protocols were co-located but not duration matched \\
SERF magnetometer \cite{Kominis2003} & $0.54\,$fT Hz$^{-1/2}$ in $0.3\,$cm$^3$ & Spectral-density operating point; interval not reported & Shielded, near-zero-field laboratory configuration & Shielding noise, vector calibration, heading error, bandwidth, and field range \\
Advanced LIGO squeezed readout \cite{Ganapathy2023,Jia2024,Capote2025LIGOO4,GWOSCO4a2025} & Quantum-noise reduction up to $5.2\,$dB at Hanford and $6.1\,$dB at Livingston; $52.6\%$ dual coincidence over the broader source interval; O4a analysis-ready fraction $0.53397$ & Detector-specific maxima and duty fractions over explicitly different intervals; not one uncertainty interval & Maintained dual-site scientific instrument with public calibrated strain & Optical loss, filter-cavity phase, classical displacement and thermal noise, calibration, and observing availability \\
Entangled-photon phase microscopy \cite{Zhang2026Imaging} & $2.76\,\mu$m resolution, phase sensitivity near $\lambda/100$, $100\,$fW illumination & Source operating-point estimates & Laboratory sample and reconstruction & Dose matching, field of view, acquisition time, scattering, detector dead time, and estimator bias \\
Rydberg RF electrometry \cite{Sedlacek2012,Holloway2014,Adams2020} & Sensitivity $\sim30\,\mu\mathrm{V\,cm^{-1}\,Hz^{-1/2}}$; minimum detected field $\sim8\,\mu\mathrm{V\,cm^{-1}}$ & Approximate source values without reported confidence interval; absolute scale follows Eq.~\eqref{eq:rydberg_field} and the projected dipole matrix element & Rb vapor cell; bright EIT resonance and Autler--Townes field inference in a laboratory microwave field & Laser linewidth, Doppler and transit-time effects, power broadening, multipath, bandwidth, and dynamic range \\
\bottomrule
\end{tblr}
\end{table}

\section{Atomic clocks and precision metrology}
\label{sec:clocks}

Frequency standards use narrow atomic, ionic, molecular, or nuclear transitions to stabilize a local oscillator and compare time or frequency across laboratories and links. Unlike a generic sensor noise floor, clock performance requires at least three separate quantities: short-term instability, systematic frequency uncertainty, and transfer or time-scale performance. Their covariance and dead time determine the uncertainty of a ratio or accumulated time error.

\subsection{Clock operation and measurement equations}

An atomic clock stabilizes a local oscillator to a transition frequency $\nu_0$. A Ramsey sequence creates a superposition, accumulates phase
\begin{equation}
\phi=2\pi(\nu_{\rm LO}-\nu_0)T,
\label{eq:ramsey}
\end{equation}
and maps the phase to a population difference \cite{Ramsey1950,Ludlow2015}. For $N$ uncorrelated atoms under projection-noise-limited white-noise conditions,
\begin{equation}
\sigma_y(\tau)\simeq\frac{1}{2\pi\nu_0CT}
\sqrt{\frac{T_c}{N\tau}},
\label{eq:clock_qpn}
\end{equation}
where $C$ is Ramsey contrast, $T_c$ the cycle time, and $\tau$ the averaging time. The expression follows by converting the binomial projection fluctuation $\Delta P\simeq(2\sqrt N)^{-1}$ at mid-fringe into phase and then averaging $\tau/T_c$ independent cycles. It excludes local-oscillator aliasing, dead time, technical detection noise, correlated atoms, and entanglement; in practice the Dick effect and transfer oscillator often determine the observed coefficient.

For contiguous nonoverlapping averages $\bar y_k(\tau)=\tau^{-1}\int_{t_k}^{t_k+\tau}y(t)\,dt$, the population Allan variance and its finite-record estimator are \cite{Allan1966,IEEE1139_2022,RileyHowe2008}
\begin{equation}
\sigma_y^2(\tau)=\frac{1}{2}\,\mathbb E\!\left[(\bar y_{k+1}-\bar y_k)^2\right],
\qquad
\widehat{\sigma}_y^2(\tau)=\frac{1}{2(M-1)}\sum_{k=1}^{M-1}(\bar y_{k+1}-\bar y_k)^2.
\label{eq:allan}
\end{equation}
Overlapping, modified, and total-deviation estimators have different transfer functions and equivalent degrees of freedom; a stability curve is physically interpretable only with its estimator, sampling, dead-time treatment, and confidence method.

Optical lattice clocks use ensembles of neutral Sr or Yb atoms trapped near a magic wavelength; single-ion clocks use species such as Al$^+$, Yb$^+$, Sr$^+$, or Hg$^+$. The apparatus comprises an ultrastable local oscillator, reference cavity, clock-interrogation optics, atom or ion trap, state readout, servo, environmental sensors, systematic-shift model, and frequency comb. Networked operation adds phase-stabilized fiber, free-space optical transfer, or satellite links.

\subsection{Demonstrated optical-clock performance}

\begin{figure}[tbp]
\centering
\includegraphics[alt={Four panels separating atomic or single-clock instability, clock-comparison instability, systematic uncertainty, and optical-frequency-ratio uncertainty.},width=0.94\linewidth]{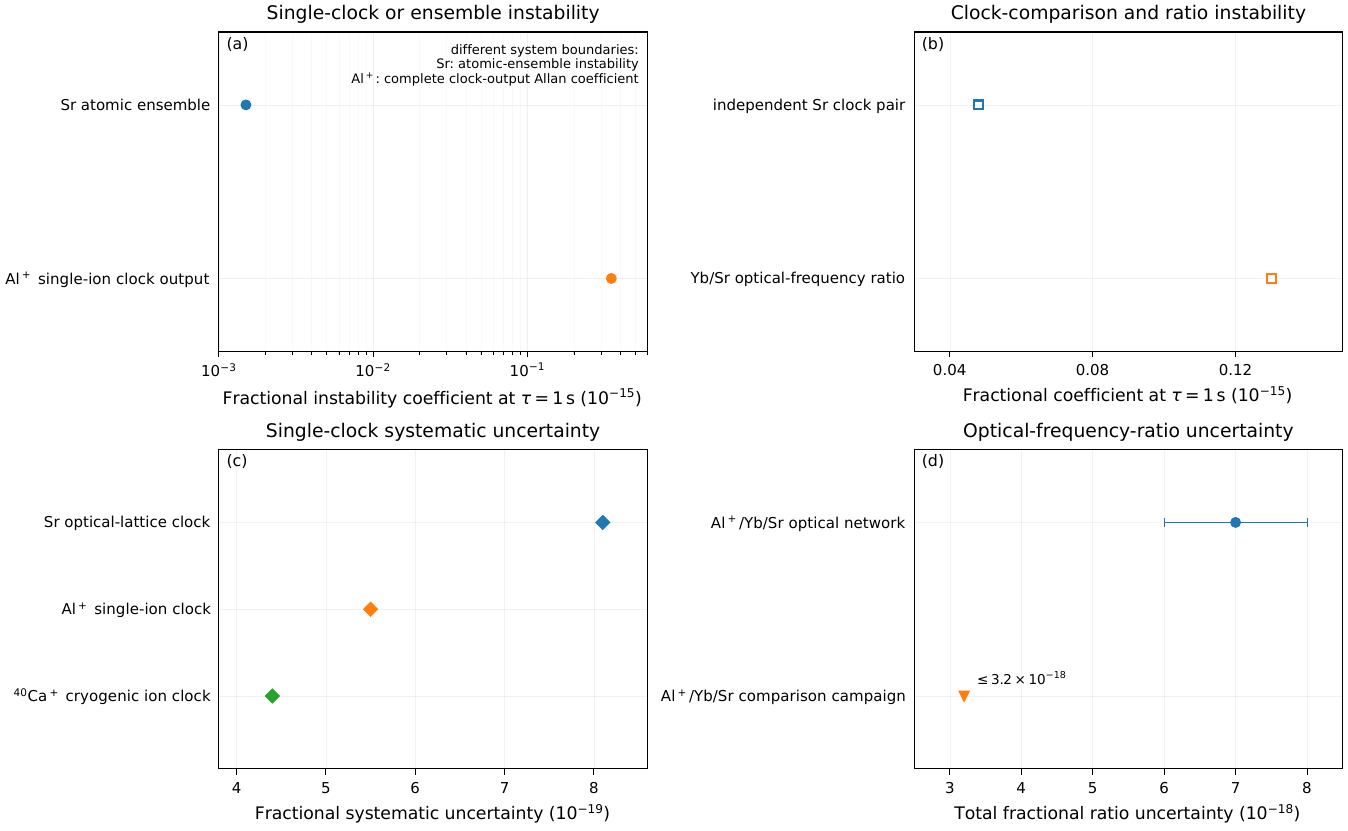}
\caption{Selected optical-frequency quantities, separated by measurand. Panels (a) and (b) plot dimensionless fractional coefficients in units of $10^{-15}$, with the exponent included in each horizontal-axis label; panel (a) uses a logarithmic scale and panel (b) a linear scale. (a) The $^{87}$Sr point is an atomic-ensemble instability coefficient, whereas the $^{27}$Al$^+$ point is the Allan-deviation coefficient of a complete single-ion clock output; the distinct system boundaries are annotated in the panel and the numerical values are not a direct ranking \cite{Kim2025Clock,Marshall2025AlClock}. (b) Independent-clock comparison and Yb/Sr ratio instability at $1\,$s \cite{Oelker2019,Aeppli2026}. (c) Evaluated single-clock fractional systematic uncertainties in units of $10^{-19}$ \cite{Zhang2026CaClock,Marshall2025AlClock,Aeppli2024}. (d) Total optical-frequency-ratio uncertainty in units of $10^{-18}$, shown as the Boulder Atomic Clock Optical Network (BACON) range and the later campaign upper bound \cite{Beloy2021,Aeppli2026}. Instability, systematic uncertainty, ranges, and upper bounds are different statistical objects and are not combined.}
\label{fig:clocks}
\end{figure}

The relevant optical-frequency quantities are different measurands and are not ranked on one axis.  The $^{87}$Sr Wannier--Stark experiment reported $118(9)\,$s coherence (source-reported fit uncertainty; coverage \NR{}) and an atomic-ensemble instability coefficient of $1.5\times10^{-18}$ at $1\,$s \cite{Kim2025Clock}.  A single-ion $^{27}$Al$^+$ clock reported instability $3.5\times10^{-16}/\sqrt{\tau/\mathrm{s}}$ and evaluated fractional systematic uncertainty $5.5\times10^{-19}$ \cite{Marshall2025AlClock}.  A liquid-nitrogen-cooled $^{40}$Ca$^+$ clock and a Sr lattice clock reported evaluated fractional systematic uncertainties of $4.4\times10^{-19}$ and $8.1\times10^{-19}$, respectively \cite{Zhang2026CaClock,Aeppli2024}.  An independent two-clock comparison reported stability $4.8\times10^{-17}$ at $1\,$s \cite{Oelker2019}.  The Boulder Atomic Clock Optical Network (BACON) reported total ratio uncertainties spanning $6$--$8\times10^{-18}$, a source-reported range rather than a confidence interval \cite{Beloy2021}; a later Al$^+$/Yb/Sr campaign reported a total ratio-uncertainty bound of $\leq3.2\times10^{-18}$ and Yb/Sr ratio instability $1.3\times10^{-16}$ at $1\,$s \cite{Aeppli2026}.  Figure~\ref{fig:clocks} separates these four measurands.

The dominant errors are blackbody-radiation shifts, Zeeman shifts, lattice and probe Stark shifts, density shifts, motional effects, cavity thermal noise, Dick-effect aliasing, dead time, frequency-comb transfer, and link phase noise.  Laboratory clock performance is compared with Cs fountains and other optical standards through traceable ratios.  A deployable timing system must additionally be compared with hydrogen masers, Global Navigation Satellite System-disciplined oscillators, and engineered holdover using time error, duty cycle, restart behavior, size, power, and maintenance interval.  A mobile ensemble of three optical clocks operated at sea for approximately three weeks, establishing autonomous and transportable operation but at stability levels substantially above the lowest-uncertainty laboratory systems \cite{Hilton2025}.

The next system-level tests are continuous comparison of independent transportable clocks, autonomous evaluation of systematic shifts, robust comb and link operation, and field time-scale generation with documented holdover and recovery.  Further reduction of intrinsic clock uncertainty has value only if transfer, uptime, and environmental corrections remain below the corresponding level.  For centimeter-scale relativistic geodesy,
\begin{equation}
\frac{\Delta\nu}{\nu}=\frac{\Delta U}{c^2}\simeq\frac{g\Delta h}{c^2},
\qquad 10^{-18}\longleftrightarrow0.92\ \mathrm{cm},
\label{eq:redshift}
\end{equation}
so the clock, transfer, local-tie, and geophysical contributions must each be controlled near the $10^{-18}$ fractional level.  Blackbody shifts provide a representative systems constraint: the mean electric-field energy scales approximately as $T^4$, so both the temperature field and the differential static and dynamic polarizabilities enter the correction and its uncertainty.  For resilient timing, the accepted output is accumulated time error,
\begin{equation}
x(t)=\int_0^t y(t')\,dt',
\label{eq:time_error}
\end{equation}
during loss of the external reference; high short-term stability does not by itself guarantee useful holdover.

\subsection{\texorpdfstring{$^{229}$Th}{229Th} nuclear-transition spectroscopy and fundamental-physics tests}

The $^{229}$Th isomer provides an optical nuclear transition near $8.4\,$eV. Radiative decay from crystal-hosted nuclei was observed before direct resonant laser excitation established precision optical access \cite{Kraemer2023Th}. Direct laser excitation of Th-doped CaF$_2$ measured $\lambda=148.3821(5)\,$nm and $\nu=2020.409(7)\,$THz, with a $630(15)\,$s fluorescence lifetime in the crystal and an inferred isolated-nucleus half-life of $1740(50)\,$s \cite{Tiedau2024Th}. A vacuum ultraviolet (VUV) frequency comb referenced to an $^{87}$Sr clock subsequently measured the nuclear-to-atomic frequency ratio and resolved the quadrupole-split structure \cite{Zhang2024ThClock}. These results establish optical access and traceable frequency comparison; they do not yet constitute a clock with a complete systematic-uncertainty budget or continuous operation. In a complementary trapped-ion route, state-selective spectroscopy of continuously supplied $^{229m}$Th$^{3+}$ measured an isolated-ion half-life of $1400^{+600}_{-300}\,$s and nuclear hyperfine constants relevant to clock-state preparation and fundamental-constant sensitivity \cite{Yamaguchi2024Th}.

The two architectures have different dominant systematics. A trapped-ion clock can use electronic transitions for cooling, state preparation, and readout, but requires production, capture, charge-state control, VUV interrogation, and ion-specific systematic evaluation. A solid-state clock can address many nuclei in the Lamb--Dicke regime, but the host introduces inhomogeneous broadening, quadrupole structure, defects, radiation damage, and temperature-dependent shifts. In CaF$_2$, the $m=\pm5/2\rightarrow\pm3/2$ line shifted $62(6)\,$kHz between $150$ and $293\,$K \cite{Higgins2025Th}. Using the central values, the mean slope over that interval is $62\,\mathrm{kHz}/143\,\mathrm{K}=0.434\,\mathrm{kHz\,K^{-1}}$; at $\nu=2.0204\times10^{15}\,$Hz, a $10^{-18}$ fractional shift then corresponds to $4.66\,\mu$K, reported here as approximately $4.7\,\mu$K. The source rounded the same requirement to $5\,\mu$K. Later measurements identified an operating temperature $196(5)\,$K where first-order thermal sensitivity vanishes and found $220\,$Hz, or $1.09\times10^{-13}$ fractional, frequency reproducibility between differently doped crystals over seven months at $195\,$K \cite{Ooi2026Th}. The same work projects sub-$10^{-18}$ temperature-shift control using quadrupole-line co-sensing; that value is a model-conditioned requirement, not a demonstrated total clock uncertainty. Laser-induced quenching has shortened the crystal isomer lifetime by a factor of three using $20\,$mW over the tested wavelength and temperature ranges, addressing cycle time at the cost of an additional optical-control channel \cite{Schaden2025Th}.

Frequency-ratio searches use the differential response
\begin{equation}
\delta\ln\!\left(\frac{\nu_i}{\nu_j}\right)=
\Delta K_{\alpha}\,\delta\ln\alpha+
\Delta K_{\mu}\,\delta\ln\mu+
\Delta K_q\,\delta\ln X_q+\cdots,
\label{eq:constant_sensitivity}
\end{equation}
where $\alpha$ is the fine-structure constant, $\mu=m_e/m_p$, $X_q$ denotes the dimensionless ratio of a light-quark mass to the quantum-chromodynamics scale, and $\Delta K$ is the difference of transition sensitivities. Optical clock--cavity and optical--microwave comparisons have constrained oscillatory ultralight scalar fields; one Yb/Cs data set spanned $298$ d with $15.4\%$ uptime \cite{Kobayashi2022DM}, atom--cavity comparisons probed masses from $10^{-21}$ to $10^{-16}\,$eV \cite{Kennedy2020DM}, and space-time-separated cavities and clocks connected by a $2220\,$km fiber link and GPS data probed $10^{-19}$ to $2\times10^{-15}\,$eV/$c^2$ \cite{Filzinger2025DM}. Transient clock networks test topological-defect signatures \cite{Derevianko2014TopologicalDM}; cavity orientation tests constrain Lorentz violation \cite{Muller2003Lorentz}; and millimeter-scale clock comparisons resolve gravitational redshift at $7.6\times10^{-21}$ measurement uncertainty \cite{Bothwell2022Redshift}. A model-dependent analysis of the nuclear transition gives $K_{\alpha}=5900(2300)$ \cite{Beeks2025ThSensitivity}, illustrating potential enhancement but also substantial nuclear-model uncertainty. For all such searches, delivered sensitivity depends on live-time duty factor, timestamp and calibration continuity, noise spectra, trial factors, and the prespecified event or spectral acceptance criterion; peak clock sensitivity alone is not the accepted discovery reach.

At operational scale, the international realization and dissemination of Coordinated Universal Time (UTC) provides the clearest example of quantum-enabled timing as a maintained service. Primary and secondary frequency standards and national UTC($k$) ensembles contribute through time-transfer links to the International Bureau of Weights and Measures (BIPM) computation of the free atomic time scale, \emph{\'Echelle Atomique Libre}, International Atomic Time (TAI), and UTC; Circular T publishes $[\mathrm{UTC}-\mathrm{UTC}(k)]$ values at five-day epochs and evaluations of the TAI scale interval \cite{BIPMTimeMetrology,BIPMCircularT}. This operational chain is distinct from the frontier optical-clock results above: its strength is traceability, continuity, institutional configuration control, and broad use rather than the lowest single-clock uncertainty.

Table~\ref{tab:clocks} separates instability, systematic uncertainty, ratio uncertainty, transportability, and nuclear-transition reproducibility. Its main conclusion is that these observables diagnose different physical limits and cannot be collapsed into one clock-performance ranking.

\begin{table}[tbp]
\caption{Selected atomic and nuclear frequency-standard configurations. Instability, systematic uncertainty, ratio uncertainty, and reproducibility are reported as distinct measurands.}
\label{tab:clocks}
\centering
\begin{tblr}{colspec={Q[l,wd={0.15\linewidth}] Q[l,wd={0.23\linewidth}] Q[l,wd={0.19\linewidth}] Q[l,wd={0.18\linewidth}] Q[l,wd={0.22\linewidth}]}, cells={valign=t}, rowsep=1.5pt, colsep=1.3pt}
\toprule
Configuration & Measurand and result & Statistical interpretation & Evidence scope & Principal limiting mechanism \\
\midrule
Sr Wannier--Stark clock \cite{Kim2025Clock} & Coherence $118(9)\,$s; ensemble instability $1.5\times10^{-18}$ at $1\,$s & Coherence fit uncertainty; instability coefficient, not accuracy & One laboratory ensemble with dynamical decoupling & Local-oscillator noise, full systematic budget, transfer, and duty factor \\
Single-clock uncertainty budgets \cite{Zhang2026CaClock,Marshall2025AlClock,Aeppli2024} & Ca$^+$: $4.4\times10^{-19}$; Al$^+$: $5.5\times10^{-19}$; Sr: $8.1\times10^{-19}$ & Source-evaluated fractional systematic uncertainties & Three species and apparatuses & Blackbody, Zeeman, Stark, density and motional corrections; independent comparison \\
Al$^+$/Yb/Sr campaign \cite{Aeppli2026} & Ratio uncertainty $\leq3.2\times10^{-18}$; Yb/Sr instability $1.3\times10^{-16}$ at $1\,$s & First value is an upper bound; second is an instability coefficient & One multispecies optical network & Long-duration independent reproduction, comb and link covariance \\
Mobile optical-clock ensemble \cite{Hilton2025} & Three-clock operation at sea for approximately three weeks & Campaign performance under transport and autonomous operation & Field demonstration & Autonomous systematic evaluation, holdover, recovery, size, power, and maintenance \\
$^{229}$Th nuclear transition \cite{Tiedau2024Th,Zhang2024ThClock,Yamaguchi2024Th,Ooi2026Th} & $148.3821(5)\,$nm and $2020.409(7)\,$THz; $220\,$Hz two-crystal reproducibility at $195\,$K over seven months & Parenthetical source uncertainties where stated; reproducibility is not a total clock uncertainty & Solid-state excitation and Sr-referenced comparison; complementary trapped-ion lifetime spectroscopy & Crystal fields, temperature shifts, radiative and internal-conversion channels, interlaboratory reproduction, and complete clock operation \\
\bottomrule
\end{tblr}
\end{table}

\section{Quantum random-number generation}
\label{sec:qrng}

Quantum random-number generators (QRNGs) convert quantum measurement outcomes into a bit stream under a declared source, detector, and adversarial model. The raw detector or digitizer stream is not the secure or validated output: digitizer response, classical and quantum side information, min-entropy estimation, extraction, online health tests, output interface, and environmental envelope determine the usable rate. Consequently, source-reported throughput is comparable only within a common trust model.

A vacuum-fluctuation integrated QRNG reported $100\,$Gbit s$^{-1}$ using a device-dependent framework that accounts for classical and quantum side information and digitizer nonlinearity \cite{Bruynsteen2023}. Frequency-domain processing in four nonoverlapping bands produced $44\,$Gbit s$^{-1}$ digitally and $52\,$Gbit s$^{-1}$ with analog processing in a 2026 experiment \cite{Li2026QRNG}. A source-device-independent optical implementation reported secure generation above $1.7\,$Gbit s$^{-1}$ \cite{Marangon2017QRNG}. A self-contained integrated photonic core delivered $2\,$Gbit s$^{-1}$; eight boards were manufactured, five were operated simultaneously for one week, and one board was monitored through $2.9$ million histograms over a 38-day QKD deployment \cite{Marangon2024QRNG}. A system-in-package implementation reported $5.2\,$Mbit s$^{-1}$ and characterization from $-40$ to $85\,^{\circ}$C \cite{Qiao2025QRNG}. These rates span different security assumptions and cannot be ordered solely by bit rate.

National Institute of Standards and Technology (NIST) entropy certificate E63 covers the named ID Quantique Quantis independent-and-identically-distributed (IID) QRNG implementations IDQ250C2, IDQ250C3, IDQ6MC1, IDQ20MC1, IDQ20MC1-S1, and IDQ20MC1-S3 under NIST Special Publication (SP) 800-90B, with a physical noise source, sample size 2, and validated entropy per sample 1.75 in the listed operating environments \cite{NISTEntropyE63,NISTSP80090B}. The certificate is evidence for those entropy-source implementations and conditions; it does not validate the complete device, its deployment history, side-channel resistance, or every downstream use. Device-independent Bell-test protocols seek stronger certification while making weaker trust assumptions about the internal device models. They impose stronger experimental requirements: a loophole-free Bell violation, sufficiently high overall detection efficiency, independent and secure setting choices, spacelike separation when a no-signaling argument is invoked, isolation against unintended communication, and finite-statistics entropy analysis \cite{Bierhorst2018}. Self-testing prepare-and-measure protocols retain explicit trusted-but-imperfect-device assumptions \cite{Lunghi2015}, whereas the cited trapped-ion certified-randomness experiment relies on a restricted adversary and computational-complexity assumptions rather than full Bell-device-independent certification \cite{Liu2025CertifiedRandomness}. In all cases, the initial seed, laboratory isolation, adversarial model, and extractor soundness remain part of the security statement.

Table~\ref{tab:qrng} organizes QRNG throughput by source model, adversarial assumption, extraction, and integration evidence. It shows that the entropy model and assurance boundary, rather than raw bit rate alone, determine which outputs are commensurate.

\begin{table}[tbp]
\caption{Selected quantum random-number generators. Output rates are comparable only within a stated source and adversarial model; certification applies only to the named implementation and conditions.}
\label{tab:qrng}
\centering
\begin{tblr}{colspec={Q[l,wd={0.18\linewidth}] Q[l,wd={0.20\linewidth}] Q[l,wd={0.20\linewidth}] Q[l,wd={0.18\linewidth}] Q[l,wd={0.22\linewidth}]}, cells={valign=t}, rowsep=1.4pt, colsep=1.0pt}
\toprule
Configuration & Extracted or validated output & Security model and processing & Integration evidence & Principal limitation \\
\midrule
Integrated vacuum-fluctuation QRNG \cite{Bruynsteen2023} & $100\,$Gbit s$^{-1}$ & Device-dependent model including classical and quantum side information and digitizer nonlinearity & Integrated high-rate laboratory implementation & Environmental robustness, independent reproduction, side-channel boundary, and sustained service \\
Frequency-multiplexed QRNG \cite{Li2026QRNG} & $44\,$Gbit s$^{-1}$ digital; $52\,$Gbit s$^{-1}$ analog & Device-dependent frequency-domain extraction in four nonoverlapping bands & Laboratory system & Common adversarial model, extractor implementation, and long-duration operation \\
Source-device-independent QRNG \cite{Marangon2017QRNG} & Secure rate above $1.7\,$Gbit s$^{-1}$ & Source-device-independent optical model & Laboratory optical implementation & Detector model, calibration, integration, and deployment evidence \\
Integrated photonic core \cite{Marangon2024QRNG} & $2\,$Gbit s$^{-1}$; eight boards; week-long five-board run; 38-day deployment record & Device-dependent source model and extractor analysis & Multi-board hardware and duration evidence & Independent multisite replication, standardized assurance, power, and service-life distribution \\
System-in-package QRNG \cite{Qiao2025QRNG} & $5.2\,$Mbit s$^{-1}$; characterized from $-40$ to $85\,^{\circ}$C & Implementation-specific entropy model & Packaged environmental test & Entropy-model robustness, side channels, population yield, and service history \\
NIST certificate E63 \cite{NISTEntropyE63,NISTSP80090B} & Validated entropy per sample 1.75 for sample size 2 for listed Quantis IID versions & SP~800-90B entropy-source validation in specified environments & Named certified implementations & Does not establish complete-device lifecycle, deployment availability, or downstream system security \\
\bottomrule
\end{tblr}
\end{table}

The cited studies establish high laboratory bit rates, integrated subsystem operation, and, for named implementations, entropy-source validation or multi-week deployment records. They do not collectively establish continuous field availability, maintenance response, failure-rate distributions, or service-level reliability for QRNGs as a general technology class. High-rate generation, statistical testing, quantum-origin certification, integration, field deployment, and maintained long-duration service are therefore treated as distinct evidence levels.

\section{Enabling technologies for quantum systems}
\label{sec:enabling}

The performance of a quantum device is inseparable from the materials, fabrication, photonic or microwave interfaces, cryogenics, vacuum, lasers, detectors, control electronics, calibration, software, packaging, and maintenance that preserve the relevant state and measurement chain. The required quantities are population distributions and integrated subsystem metrics, not only selected-device records.
\subsection{Materials loss, coherence, and process distributions}

Complementary metal--oxide--semiconductor (CMOS)-compatible processing is relevant to semiconductor and superconducting platforms, but cryogenic quantum-device yield depends on low-temperature distributions rather than room-temperature process compliance alone. Quantum-device performance is commonly limited by interfaces and defects rather than by the intended Hamiltonian.  For superconducting resonators and qubits, a participation model is
\begin{equation}
\frac{1}{Q_i}\simeq\sum_j p_j\tan\delta_j
+\frac{1}{Q_{\rm rad}}+\frac{1}{Q_{\rm qp}}+\frac{1}{Q_{\rm seam}}+\cdots,
\label{eq:participation}
\end{equation}
where $p_j$ is the electric-field participation of material region $j$ and $\tan\delta_j$ its dielectric loss tangent.  Surface oxides, processing residue, junction barriers, substrate interfaces, seam currents, radiation, magnetic flux, and nonequilibrium quasiparticles generate device-to-device variation.  Tantalum-based transmons have demonstrated coherence exceeding $0.3\,$ms in selected devices \cite{Place2021}; the manufacturing observable is the joint distribution of $T_1$, $T_2$, frequency, anharmonicity, gate error, and package loss across sites, wafers, and cooldowns. A 300-mm CMOS-compatible superconducting-qubit study reported 393 functional qubits among 400 tested and $T_1$ values spanning approximately $42$--$113\,\mu$s across the reported wafer data, illustrating the distinction between selected-device records and manufacturing distributions \cite{VanDamme2024}.

For semiconductor spins, isotopic enrichment suppresses nuclear-spin noise while interface roughness, oxide traps, charge noise, valley splitting, gate-stack variation, and tunnel-coupling dispersion determine tuning range and exchange control. A 300-mm study covering 232 twelve-quantum-dot devices reported $100\%$ contact and gate yield, $3703/3712=99.8\%$ quantum-dot formation yield, $223/232$ full-device yield (reported as $96\%$), and a threshold-voltage random variation of $59\,$mV standard deviation \cite{Neyens2024}.  For color centers, the relevant distributions are position, charge-state stability, strain, optical linewidth, spectral diffusion, spin coherence, and nanophotonic coupling.  Atomic platforms exchange solid-state disorder for vacuum, laser, magnetic, electric-field, and collision constraints; collision-limited lifetime and stray-field gradients must be characterized over the full operating volume.

\subsection{Integrated photonics, sources, detectors, and transduction}

An integrated quantum optical path contains source generation, collection, propagation, splitting, switching, interference, conversion, detection, and often cryogenic packaging.  The end-to-end probability is the product of the unconditional efficiencies.  A manufacturable silicon-photonic platform reported multimode and single-mode silicon-nitride (SiN) propagation losses of $0.5\pm0.3$ and $1.8\pm0.2\,$dB m$^{-1}$, splitter loss $0.5\pm0.2\,$mdB, crossing loss $1.2\pm0.4\,$mdB, and fiber--chip coupling loss $52\pm12\,$mdB in the stated process \cite{Alexander2025}.  These source-reported intervals are retained without an inferred confidence level or coverage factor (\NR{}).  The same work reported a $98.9\%$ median \emph{on-chip detection efficiency} for its best-performing four- and five-cell waveguide-integrated photon-number-resolving detectors; this boundary excludes fiber coupling and upstream optical-path loss. Conditional module fidelities are listed in Table~\ref{tab:computing}.  Conditional fidelity and unconditional success probability must be propagated separately; a high-fidelity postselected event can coexist with an inadequate event rate.

Single-photon detectors are further characterized by dark-count rate, timing jitter, dead time, saturation, photon-number resolution, crosstalk, and array yield.  Microwave--optical and other quantum transducers require bidirectional efficiency, added noise referred to the input, bandwidth, pump isolation, and preservation of entanglement or state fidelity.  A conversion efficiency without an added-noise and bandwidth measurement is not a quantum interconnect specification.

\subsection{Cryogenics, vacuum, lasers, and classical control}

At microwave frequencies, Eq.~\eqref{eq:thermal_occupation} makes thermalization a first-order error source.  Superconducting qubits, some spin qubits, superconducting nanowire detectors, transition-edge sensors, and parametric amplifiers require cryogenic operation.  The relevant system quantities are device temperature rather than refrigerator base temperature, cooling power at each stage, conductive and radiative heat load, attenuation and filtering, amplifier back-action, vibration, magnetic field, cooldown interval, and service availability.  Cryogenic multiplexing and low-power cryo-CMOS reduce room-temperature wiring but introduce dissipation, electromagnetic coupling, and calibration complexity \cite{Reilly2015,Bardin2021}.

Atomic clocks, trapped ions, neutral atoms, interferometers, and Rydberg sensors require laser linewidth, phase and intensity noise, pointing, switching time, optical power, and long-term lock robustness to be specified at the atoms.  Vacuum qualification includes pressure or collision lifetime, outgassing, getter or pump capacity, optical contamination, thermal cycling, shock and vibration, and recovery from loss of control.  State readout and feedback must be synchronized to the physical dynamics; estimator or decoder throughput, queue stability, and calibration age are system measurands rather than software implementation details. A distance-5 surface-code memory on a 72-qubit superconducting processor sustained real-time decoding for as many as $10^6$ syndrome cycles, with $1.1\,\mu$s cycle time, net average decoder latency $63\pm17\,\mu$s, and throughput below $1.1\,\mu$s per cycle \cite{Acharya2025}. The source estimated data-transfer latency below $10\,\mu$s and had not yet implemented feedback, so the measured decoder throughput closes only part of the complete reaction-time budget.

\subsection{Functional yield, serial interfaces, and calibration scaling}

If $M$ statistically independent elements are all required and element $i$ has yield $y_i$, the series-system yield is $Y=\prod_i y_i$; for identical elements,
\begin{equation}
Y=y^M,
\qquad
\frac{u_Y}{Y}\simeq M\frac{u_y}{y}
\label{eq:yield}
\end{equation}
under first-order propagation of a common estimated $y$. The model is exact only for independent, indispensable elements with a fixed pass/fail criterion. At $y=0.999$, $Y=0.368$ for $M=1000$; conversely, $Y\geq0.90$ requires $y\geq0.999895$. Correlated lithographic defects can make the tail worse than Eq.~\eqref{eq:yield}, whereas repair, redundancy, and binning change the system architecture. The physical measurement needed for scaling is therefore the joint lot-level yield and failure-correlation structure, not only the mean device pass fraction.

Serial loss gives a similarly direct architecture constraint. For identical interfaces with power or success loss $\ell$ in decibels, Eq.~\eqref{eq:serial_loss} becomes $\eta_{\rm sys}=10^{-M\ell/10}$, so retaining at least a fraction $\eta_{\min}$ requires
\begin{equation}
\ell\leq-\frac{10}{M}\log_{10}\eta_{\min}.
\label{eq:interface_requirement}
\end{equation}
For $M=100$ and $\eta_{\min}=0.5$, $\ell\leq0.030\,$dB per interface. The relation assumes a cascade without gain, heralded bypass, or multiplexed rerouting. A common coupling offset accumulates coherently in decibels, while independent calibration errors add in quadrature. Consequently, a quoted component loss must be accompanied by package, thermal-cycle, wavelength, polarization, and path-length distributions before it can support a system transmission claim.

Calibration also scales with architecture.  An unstructured all-pairs interaction graph has $N(N-1)/2$ pair parameters; local or modular control reduces this number but introduces hierarchy and interface parameters.  Embedded monitors, symmetry, physics-informed models, and automated calibration are required to keep the calibration state current relative to drift.

\begin{figure}[tbp]
\centering
\includegraphics[alt={Three analytical panels showing multiplicative assembly yield, serial-interface transmission, and calibration-parameter scaling.},width=0.96\linewidth]{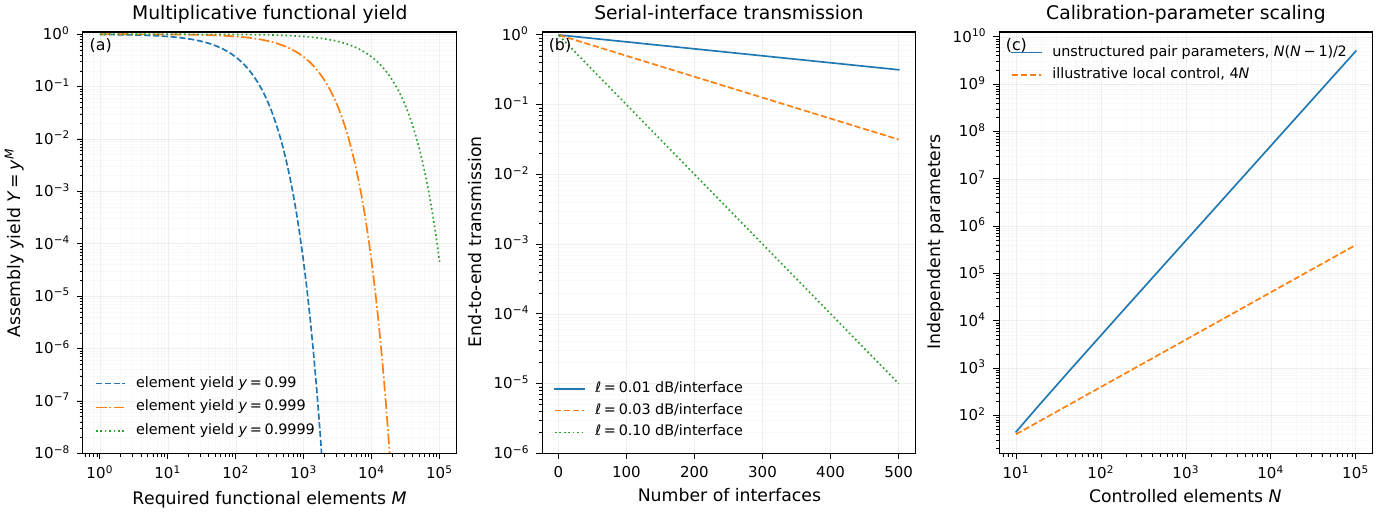}
\caption{Scale-dependent engineering constraints evaluated from explicit models. (a) Assembly yield $Y=y^M$ for independent required elements; correlated defects and repairable redundancy are omitted. (b) End-to-end transmission through identical serial interfaces. (c) Calibration-parameter scaling for an unstructured all-pairs graph and an illustrative local-control model. These analytical panels use common denominators and replace a mixed-percentage comparison of unrelated manufacturing quantities. Source-reported population evidence, including $393/400$ functional superconducting qubits and $223/232$ twelve-dot spin arrays passing the full-device criterion (reported as $96\%$), is summarized in Table~\ref{tab:infrastructure}.}
\label{fig:manufacturing}
\end{figure}

Figure~\ref{fig:manufacturing} shows why device distributions, serial loss, and calibration scaling matter before a component record can support a large system. Table~\ref{tab:infrastructure} combines selected component results with wafer- and device-population evidence.

\begin{table}[tbp]
\caption{Selected quantitative enabling-infrastructure operating points and required scale-up measurements; row order is nonranking. Source-reported ``$\pm$'' intervals are retained; coverage is \NR{} where the accessible primary text does not specify it. Unless identified as analytical, numerical entries are source-transcribed.}
\label{tab:infrastructure}
\centering
\begin{tblr}{colspec={Q[l,wd={0.13\linewidth}] Q[l,wd={0.21\linewidth}] Q[l,wd={0.17\linewidth}] Q[l,wd={0.21\linewidth}] Q[l,wd={0.20\linewidth}]}, cells={valign=t}, rowsep=1.7pt, colsep=1.5pt}
\toprule
Subsystem / configuration & Direct measurand and value & Operating condition & Integration significance & Scale-up qualification evidence \\
\midrule
Tantalum transmon devices \cite{Place2021} & Coherence $>0.3\,$ms in selected devices & Millikelvin superconducting circuit & Selected device-level materials result & Wafer and package distributions, cooldown repeatability, gate and correlated-error statistics \\
300-mm superconducting-qubit population \cite{VanDamme2024} & $393/400=98.25\%$ functional, fully characterized qubits; time-averaged $T_1$ range $42$--$113\,\mu$s, median $75\,\mu$s & One 300-mm wafer; 400 qubits on 32 D1 and 24 D2 subdies, measured at $10\,$mK without prescreening & Wafer-scale population evidence from one fabrication run & Independent lots, packaged yield, cooldown repeatability, frequency-collision and gate-error distributions \\
300-mm silicon-spin device population \cite{Neyens2024} & $223/232$ twelve-dot arrays passed the source-defined full-device criterion (reported as $96\%$); contact and gate yield $100\%$, quantum-dot yield $3703/3712=99.8\%$ & 232 devices across 58 dies, four devices per die, measured with a $1.6\pm0.2\,$K cryogenic wafer prober; full-device pass requires all contacts, gates, and quantum dots to yield & Wafer-scale transport and electrostatic evidence from one representative wafer & Millikelvin qubit-operation yield, coupled-array tuning, packaged reliability, and independent lot replication \\
SiN passive photonics \cite{Alexander2025} & Propagation $0.5\pm0.3$ and $1.8\pm0.2\,$dB m$^{-1}$; splitter $0.5\pm0.2\,$mdB; crossing $1.2\pm0.4\,$mdB; fiber coupling $52\pm12\,$mdB & Foundry-compatible silicon photonics; source-reported intervals, coverage \NR{} & Component and interface loss budget & Wafer maps, path-length and channel-count scaling, thermal cycling and packaged loss \\
Photon-number-resolving detector \cite{Alexander2025} & Median on-chip detection efficiency $98.9\%$ for best- performing 4- and 5-cell waveguide-integrated devices; not an end-to-end system efficiency & Cryogenic integrated detector; optical coupling and upstream path excluded & Readout component & Array uniformity, dark counts, jitter, saturation, crosstalk and service interval \\
300-mm silicon-spin unit cell \cite{Steinacker2025} & All reported operations $>99\%$ & Foundry-compatible quantum-dot device & Manufacturing-compatible component & Wafer distributions, automated tuning, cryogenic electronics and coupled-array yield \\
Microwave mode at $5\,$GHz & $\bar n_{\rm th}\simeq6\times10^{-6}$ at $20\,$mK from Eq.~\eqref{eq:thermal_occupation} & Ideal thermal equilibrium & Analytical thermal requirement & Device temperature, nonequilibrium photons, filtering, infrared load and quasiparticles \\
Real-time surface-code decoding \cite{Acharya2025} & Net average decoder latency $63\pm17\,\mu$s; throughput-equivalent processing time below $1.1\,\mu$s per cycle for runs up to $10^6$ cycles & Distance-5 code on a 72-qubit processor; $1.1\,\mu$s cycle; source interval is one standard deviation & Real-time decoder sustained below-threshold memory operation; data transfer estimated $<10\,\mu$s and feedback not implemented & Reaction latency and memory scaling with code size; queue growth, hardware transfer, feedback integration, drift, and fault recovery \\
\bottomrule
\end{tblr}
\end{table}
\subsection{Technical standards, assurance, and interoperability}
\label{sec:standards}

Technical standards are relevant when the reported output depends on common definitions, traceability, interface behavior, or implementation assurance. NIST SP 800-90B specifies entropy-source design, health tests, and min-entropy validation \cite{NISTSP80090B}; certificate E63 applies only to the named Quantis IID implementations and listed environments, not to complete-device lifecycle or side-channel assurance \cite{NISTEntropyE63}. ETSI GS QKD 016 defines Common Criteria requirements for a QKD module through final secret-key output but does not establish network availability or endpoint security \cite{ETSIQKD016}. The International Organization for Standardization/International Electrotechnical Commission (ISO/IEC) Joint Technical Committee (JTC) 3 provides the broader international venue for quantum-technology terminology, metrology, characterization, interfaces, and test methods \cite{ISOJTC3_2024}.

\subsection{AI-assisted calibration, control, characterization, and design}
\label{sec:ai}

Machine-learning methods are relevant when they reduce the physical cost of calibration, improve a same-device quantum operation, infer a model or state with quantified uncertainty, or satisfy a real-time control deadline. For supervised calibration with features $\boldsymbol x_i$, targets $\boldsymbol y_i$, and model $f_{\boldsymbol\theta}$,
\begin{equation}
\boldsymbol\theta^\star=\arg\min_{\boldsymbol\theta\in\Theta}
\left[\frac{1}{N}\sum_{i=1}^N \ell\!\left(f_{\boldsymbol\theta}(\boldsymbol x_i),\boldsymbol y_i\right)
+\lambda\mathcal R(\boldsymbol\theta)\right],
\label{eq:supervised_calibration}
\end{equation}
where the feasible set $\Theta$ must encode hardware-safe voltage, optical-power, microwave-power, and slew-rate bounds. A learned calibration is validated on held-out devices, dates, or drift states, not only on randomly held-out samples from one acquisition.

\begin{figure}[tbp]
\centering
\includegraphics[alt={Six panels containing only commensurate within-panel quantities: quantum-dot tuning time, transmon leakage probability, neutral-atom assignment-error reduction, inference timing, AlphaQubit data accounting, and matched-filter readout acceleration.},width=0.98\linewidth]{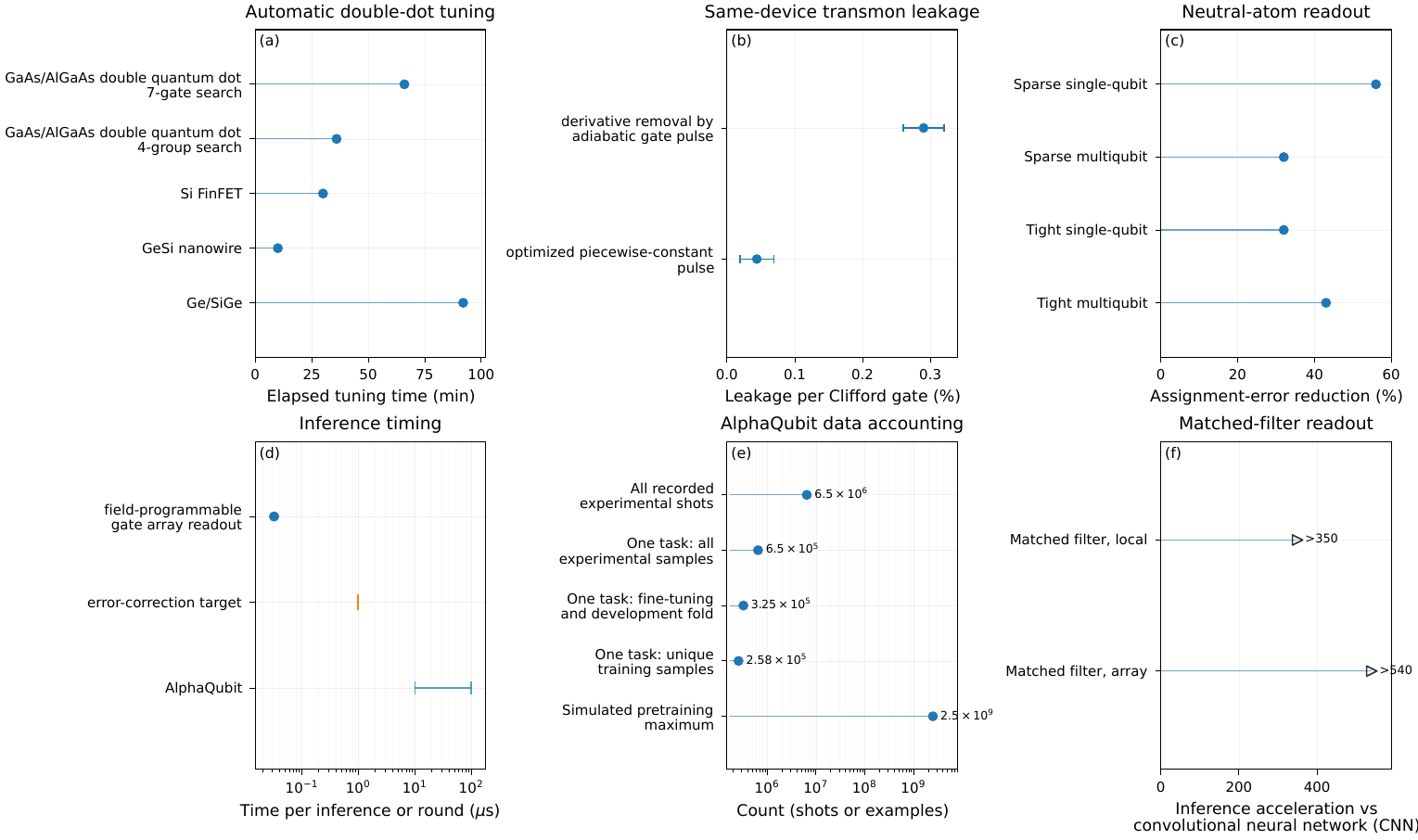}
\caption{Commensurate quantitative comparisons for demonstrated AI-assisted quantum engineering. (a) Elapsed automatic double-dot tuning times in minutes \cite{Moon2020,Severin2024}. The first two entries are GaAs/AlGaAs gate-defined double quantum dots labeled by the active search space (seven gates for device 2 and four grouped controls for device 1); the remaining entries are labeled by device architecture. (b) Direct leakage-randomized-benchmarking estimates for same-device $4.16\,$ns derivative removal by adiabatic gate (DRAG) and optimized piecewise-constant (PWC) transmon pulses: $0.29(3)\%$ and $0.044(25)\%$, respectively; the capped intervals are source-reported one-standard-deviation fit uncertainties \cite{Werninghaus2021}. (c) Relative reduction in neutral-atom assignment error against Gaussian thresholding; each point retains its sparse/tight and single-/multiqubit denominator, while the distinct crosstalk-correlation reduction remains in Table~\ref{tab:ai_cases} \cite{Phuttitarn2024}. (d) Inference timing: the FPGA point is a source-reported deterministic single-shot readout latency, the $1\,\mu$s marker is the superconducting syndrome-cycle target, and the horizontal AlphaQubit segment is the source-reported $10$--$100\,\mu$s computation-time interval per decoding round with no reported central estimate; it excludes final-answer latency after receipt of the last syndrome round \cite{Bausch2024,DiGuglielmo2025}. (e) AlphaQubit data accounting. The complete experimental corpus contains $6.5\times10^6$ recorded shots; one distance--basis--location task contains $650{,}000$ samples, of which $325{,}000$ form the fine-tuning/development fold, $258{,}440$ are unique training samples, and $66{,}560$ are development samples. The other $325{,}000$ samples are held out for final testing. Simulated pretraining used up to $2.5\times10^9$ examples; repeated passes through a fixed experimental set are not counted as new examples \cite{Bausch2024}. (f) Source-reported lower bounds on matched-filter inference acceleration relative to the corresponding convolutional neural network (CNN) baselines \cite{Kent2026Matched}. Filled circles denote point-valued measurements, source-reported counts, or explicitly derived counts. Only panel (b) shows statistical uncertainty intervals; panel (d) shows a reported operating-time range without a central estimate, panel (e) shows counts, and open triangles in panel (f) denote lower bounds. Comparisons are valid only within a panel.}
\label{fig:ml_quantum}
\end{figure}

For coherent control of a $d$-dimensional computational subspace, a standard unitary objective is
\begin{equation}
F_{\rm pro}(\boldsymbol\theta)=\frac{1}{d^2}
\left|\operatorname{Tr}\!\left(U_{\rm target}^{\dagger}U_{\boldsymbol\theta}\right)\right|^2,
\label{eq:pulse_fidelity}
\end{equation}
augmented in practice by penalties for leakage, duration, amplitude, bandwidth, and heating. Hamiltonian and noise learning use likelihood-based inference,
\begin{equation}
p(\boldsymbol\lambda\mid\mathcal D)\propto
p(\mathcal D\mid\boldsymbol\lambda)p(\boldsymbol\lambda),
\label{eq:bayes_hamiltonian}
\end{equation}
but posterior concentration within an incorrect model class is not a physical validation. State reconstruction is a constrained estimation problem,
\begin{equation}
\widehat\rho=\arg\min_{\rho\succeq0,\,\operatorname{Tr}\rho=1}
\Big[-\sum_k n_k\log\operatorname{Tr}(E_k\rho)+\lambda\mathcal R(\rho)\Big].
\label{eq:state_reconstruction}
\end{equation}
Adaptive sensing chooses the next action by expected information gain, while drift compensation is sequential state estimation. In both cases the estimator latency, uncertainty calibration, rejected measurements, and control-induced dead time enter the instrument performance. The semiconductor examples below use double quantum dot devices; the pulse comparison uses derivative removal by adiabatic gate (DRAG) and optimized piecewise-constant (PWC) waveforms; the readout implementation uses a radio-frequency system-on-chip (RFSoC) incorporating a field-programmable gate array (FPGA); neutral-atom readout is compared using convolutional neural networks (CNNs) and matched filters; and compiled-gate error is evaluated with interleaved randomized benchmarking (IRB). Figure~\ref{fig:ml_quantum} and Table~\ref{tab:ai_cases} summarize the demonstrated hardware cases: the figure separates commensurate within-panel observables, while the table retains each task, baseline, operating condition, and unreported cost term. Their main conclusion is that measured improvements are task specific; none alone establishes cross-device transfer or an end-to-end autonomous-system gain.

%\Needspace{0.35\textheight}
\begin{longtblr}[
caption={Representative AI-assisted quantum-engineering experiments. Improvements are reported only relative to the stated same-task baseline; training and inference costs are included where the source reports them.},
label={tab:ai_cases}
]{colspec={Q[l,wd={0.14\linewidth}] Q[l,wd={0.19\linewidth}] Q[l,wd={0.18\linewidth}] Q[l,wd={0.25\linewidth}] Q[l,wd={0.18\linewidth}]}, cells={valign=t}, rowsep=1.3pt, colsep=2.0pt}
\toprule
Platform and task & Model and data & Baseline & Quantitative result & Principal limitation \\
\midrule
GaAs double-dot tuning \cite{Moon2020} & Gaussian-process search and learned classifiers in up to eight gate voltages & Human benchmark and pure random search without peak detection & Median complete tuning below $70\,$min; grouped-control run $36\,$min; source-inferred $179\times$ pure-random-search improvement & $179\times$ is an ablation-model comparison, not a net gain including training, human intervention, and rejected runs \\
Cross-architecture quantum-dot tuning \cite{Severin2024} & One algorithm on Si FinFET, GeSi nanowire, and Ge/SiGe devices & Random search under architecture-specific bounds & $30$, $10$, and $92\,$min tuning times & Architecture-specific safe bounds and small population \\
Transmon pulse optimization \cite{Werninghaus2021} & Closed-loop sample-level waveform optimization & Same-device $4.16\,$ns derivative removal by adiabatic gate (DRAG) pulse & Optimized leakage $0.044(25)\%$ versus directly measured DRAG leakage $0.29(3)\%$; central-value ratio $6.6$; Clifford fidelity $99.76(8)\%$ versus $99.11(8)\%$ & Up to $25\,$h optimization for the largest tested parameterization; transfer to large arrays \\
Cross-resonance and compiled SWAP \cite{Baum2021} & Experimental deep reinforcement learning on superconducting qubits & Hardware-default gates and black-box optimizers on the same device & Cross-resonance gate repetition: $F_0=98.95(7)\%$, $F_1=99.560(58)\%$, $\epsilon_0/\epsilon_1=2.39(35)$, and $\Delta F=0.610(91)$ percentage point; compiled-SWAP interleaved randomized benchmarking (IRB): $F_0=97.68(12)\%$, $F_1=98.40(12)\%$, $\epsilon_0/\epsilon_1=1.45(13)$, and $\Delta F=0.72(17)$ percentage point & Ratios are review-derived from $\epsilon=1-F$ with independent fit-error propagation; hardware-learning cost and transfer to other devices, couplers, and drift processes \\
Adaptive relaxation monitoring \cite{Berritta2026} & FPGA Bayesian tracking of two transmons & Nonadaptive acquisition & Millisecond-scale estimates; approximately $10^2$ improvement in temporal resolution; switching to $10\,$Hz & Millisecond tracking demonstrated; autonomous scheduling and logical-error response remain to be measured \\
Neutral-atom readout \cite{Phuttitarn2024,Kent2026Matched} & Single- and multiqubit CNNs; local and array matched filters & Gaussian thresholding and corresponding CNNs & Sparse-array assignment-error reductions $56\%/32\%$ and tight-array reductions $32\%/43\%$ (single/multiqubit); crosstalk-correlation reduction $78.5\%$; matched filters $>350\times/>540\times$ faster & Assignment error, correlation, and acceleration use distinct denominators; cross-device transfer and full-array feedback deadline not reported \\
FPGA transmon readout \cite{DiGuglielmo2025} & Quantization-aware neural network in an RFSoC FPGA; $5\times10^5$ shots per prepared state & Threshold and matched-filter discriminators & $96\%$ single-shot fidelity, $32.25\,$ns latency, $<16\%$ lookup-table use & Accuracy comparable to simple baselines for a single qubit \\
\bottomrule
\end{longtblr}

AlphaQubit is reported quantitatively in Table~\ref{tab:ai_decoders} rather than duplicated in Table~\ref{tab:ai_cases}. On superconducting hardware, closed-loop optimization of a $4.16\,$ns pulse reduced directly measured leakage from $0.29(3)\%$ for the same-duration DRAG calibration to $0.044(25)\%$, a central-value ratio of $6.6$, while Clifford fidelity increased from $99.11(8)\%$ to $99.76(8)\%$ \cite{Werninghaus2021}. For the deep-reinforcement-learning gate study, define infidelity $\epsilon=1-F$. Cross-resonance gate repetition gave $F_0=98.95(7)\%$ and $F_1=99.560(58)\%$, hence $\epsilon_0/\epsilon_1=2.39(35)$, a relative infidelity reduction of $58.1(6.2)\%$, and an absolute fidelity increase $F_1-F_0=0.610(91)$ percentage point. Compiled-SWAP interleaved randomized benchmarking gave $F_0=97.68(12)\%$ and $F_1=98.40(12)\%$, hence $\epsilon_0/\epsilon_1=1.45(13)$, a relative infidelity reduction of $31.0(6.3)\%$, and an absolute fidelity increase $0.72(17)$ percentage point; the propagated ratio uncertainties treat the source fit errors as independent \cite{Baum2021}. The same study reported calibration-free robustness for as long as 25 days. The automatic quantum-dot-tuning system achieved median tuning below $70\,$min and a source-inferred $179\times$ advantage over pure random search without peak detection; that ratio is an ablation-model comparison rather than a net wall-time gain including training and intervention \cite{Moon2020}. 

In neutral-atom readout, CNNs reduced assignment error by $56\%$ and $32\%$ in sparse arrays and by $32\%$ and $43\%$ in tightly spaced arrays for single- and multiqubit models, respectively; the multiqubit model reduced nearest-neighbor crosstalk correlation by $78.5\%$ \cite{Phuttitarn2024}. Physically interpretable matched filters retained maximum error reductions of $32\%$ and $43\%$ and were more than $350\times$ and $540\times$ faster than the corresponding local and array CNNs \cite{Kent2026Matched}. An RFSoC neural discriminator achieved $32.25\,$ns deterministic inference with $96\%$ single-shot transmon-readout fidelity, while the simpler single-qubit baselines reached comparable classification accuracy \cite{DiGuglielmo2025}. An FPGA Bayesian monitor tracked relaxation fluctuations in two transmons within a few milliseconds, improving temporal resolution by approximately two orders of magnitude and resolving switching rates up to $10\,$Hz; the experiment demonstrated monitoring rather than closed-loop improvement of logical error or availability \cite{Berritta2026}.

The dominant limitations are distribution shift, synthetic-to-experimental mismatch, exploration cost, weak baselines, lack of calibrated uncertainty, and deterministic latency. An AI controller must include bounded actions, watchdogs, a physics-based fallback, and a retraining policy that does not invalidate the quoted error or uncertainty. Generalization across nominally identical devices is not guaranteed because fabrication and environmental variation can exceed the training distribution.

%\FloatBarrier
\section{Synthesis and outlook}
\label{sec:limits}
\label{sec:conclusion}

The principal cross-domain result is that performance is not invariant under a change of denominator. Moving from physical-operation error to logical-cycle error, conditional fidelity to unconditional success, active time to elapsed time, peak sensitivity to analysis-ready operation, or an ideal classical comparator to a device-matched one can move or reverse a conclusion without changing the underlying quantum hardware. Denominator dependence is therefore not a reporting detail. It identifies the system boundary that must be engineered, measured, and validated before a component record can support an advantage claim.

The field is governed by a small set of mechanisms whose consequences differ by function. No-cloning and measurement backaction constrain communication, verification, and error-correction architectures. Loss and finite detection efficiency set exponential or multiplicative penalties in photonic links and erase metrological squeezing. Thermal and nonequilibrium occupation limit microwave devices and transducers. Correlated noise and calibration covariance determine whether component errors remain correctable when channel and system size increase. These constraints are physical; materials, encoding, cooling, multiplexing, fabrication, control, and maintenance determine how closely an implementation approaches them.

Table~\ref{tab:physics_paths} follows the chain used throughout the review: physical mechanism, governing relation, measurable quantity, present evidence, and required advance. It is a synthesis of mechanisms rather than a maturity score.

\begin{longtblr}[
caption={Physical mechanisms that presently limit quantum technologies and the measurements required to change the assessment.},
label={tab:physics_paths}
]{colspec={Q[l,wd={0.12\linewidth}] Q[l,wd={0.17\linewidth}] Q[l,wd={0.18\linewidth}] Q[l,wd={0.26\linewidth}] Q[l,wd={0.23\linewidth}]}, cells={valign=t}, rowsep=1.3pt, colsep=1.5pt}
\toprule
Physical mechanism & Governing relation or scaling & Measurable quantity & Representative present evidence & Required advance \\
\midrule
Decoherence and correlated error & Union bound $P_{\rm fail}\leq\sum_iN_ip_i$ (no independence assumption); the rare-event approximation and use of isolated $p_i$ require limited overlap and context-matched marginal probabilities & $T_1$, $T_2$, leakage, spatial/temporal correlation spectrum, burst rate and recovery, logical error per cycle & Distance-7 surface-code memory at $(1.43\pm0.03)\times10^{-3}$ per cycle; in the cited superconducting devices and intervals, radiation-induced events produced chip-wide millisecond-scale degradation, including a cosmic-ray component at $1/(592^{+48}_{-41})\,$s in a ten-transmon array \cite{McEwen2022Cosmic,Harrington2025Cosmic} & Gap engineering, phonon/quasiparticle trapping, event tagging, erasure-aware decoding, and repeated logical workloads under device- and environment-specific nonstationary noise \\
Multiplicative loss & $\eta_{\rm sys}=\prod_i\eta_i$ & Source-to-detector probability, insertion loss, Bell-pair or secret-key rate & High conditional photonic fidelities and long-distance QKD coexist with small unconditional event rates & Integrated source, switch, conversion, memory, and detector chain whose delivered rate exceeds the best direct alternative \\
Thermal occupation and nonequilibrium excitation & $\bar n_{\rm th}=[\exp(hf/\kB T)-1]^{-1}$ & Device-mode occupation, quasiparticle density, added noise, cooling power & A $5\,$GHz equilibrium mode at $20\,$mK has $\bar n_{\rm th}\simeq6\times10^{-6}$, but device temperatures and nonequilibrium photons can be larger & In situ mode thermometry, lower-dissipation control, filtered interconnects, and reproducible cooldown behavior at system scale \\
Calibration drift and covariance & $C_\theta\simeq JC_vJ^{\mathsf T}$; $u_c/u_{\rm ind}=\sqrt{1+(M-1)\rho}$ & Calibration age, cross-covariance, change points, restart error & Optical clocks and large processors report precise point estimates, while full long-duration covariance and recovery data are uncommon & Continuous calibration with traceable uncertainty, automated change detection, and predeclared recovery criteria \\
Control and measurement bandwidth & Physical deadline set by gate, syndrome, sensor, or network cycle & Worst-case latency, queue depth, update rate, duty factor & Surface-code cycles near $1\,\mu$s coexist with a source-reported AlphaQubit computation-time interval of $10$--$100\,\mu$s per round, without a central estimate; RFSoC readout can reach tens of ns & Deterministic low-power inference and control that remains stable under peak data rate and hardware drift \\
Fabrication variability and yield & $Y=y^M$ for independent required elements & Wafer distributions, full-device yield, coupled-module yield, lifetime & $393/400$ functional superconducting qubits in one 300-mm study and $223/232$ twelve-dot spin arrays passing the full-device criterion (reported as $96\%$) & Lot-level coupled-system yield, package and thermal-cycle reliability, repair strategy, and multisource qualification \\
Measurement efficiency and backaction & $v_{\rm obs}=\eta v_{\rm in}+1-\eta$; Fisher-information loss & Detection efficiency, squeezing at estimator, readout error, demolition probability & LIGO reports up to $5.2$--$6.1\,$dB quantum-noise reduction in selected bands; photonic and spin readout remain interface limited & Higher end-to-end efficiency with stable phase, calibrated backaction, and bandwidth-wide performance \\
Memory and stochastic network waiting & $\langle t_{\rm wait}\rangle\simeq(f_{\rm rep}p_{\rm succ})^{-1}$; multihop tails set storage demand & Memory lifetime and fidelity, interface efficiency, waiting-time distribution & Remote-memory entanglement over $14.5\,$km with $78.6\%\pm2.0\%$ Bell-state fidelity & Multiplexed multi-link route with swapping and end-to-end rate/fidelity above direct transmission under matched conditions \\
\bottomrule
\end{longtblr}

Several capabilities are established at different integration levels. Atomic time standards and squeezed gravitational-wave readout operate as maintained systems, and selected QKD configurations have been fielded. QRNG studies demonstrate high laboratory bit rates, integrated multi-board operation, and named entropy-source certifications, but the cited evidence does not establish a general class of continuously maintained field services. Optical clocks have evaluated systematic uncertainties below $10^{-18}$ in several species. Quantum processors have demonstrated high-fidelity physical control, repeated logical-memory suppression, real-time decoding, and early logical operations. Programmable simulators measure many-body observables in regimes that challenge direct classical calculation, and network experiments have distributed entanglement through memories and metropolitan links.

The remaining advances are function specific. Fault-tolerant computation requires repeated logical gates, routing, non-Clifford-state production, deterministic decoding, and verification to close in one reproduced workload. Quantum simulation requires prospective predictions confirmed outside calibration data and against continuously improving tensor-network, neural-quantum-state, sparse-operator, Monte Carlo, and exact-diagonalization methods. Neural quantum states are a competitive baseline when their energy, observables, sampling cost, and optimization stability are demonstrated at the same Hamiltonian, size, symmetry sector, and error target rather than inferred from representational capacity alone \cite{Carleo2017NQS,Sharir2020NQS,HibatAllah2020NQS,Roth2023NQS,Szabo2020NQSSign}. Communication requires multiplexed memories and repeater routes that outperform direct transmission at equal fidelity, security, and elapsed time. Sensors and clocks require bandwidth, bias, transfer, covariance, uptime, and recovery to remain controlled outside the laboratory. Enabling hardware must progress from selected devices to lot-level distributions of yield, lifetime, thermal load, and repair.

AI-assisted tuning, pulse optimization, readout, Hamiltonian identification, decoding, and drift tracking can accelerate this engineering loop when the comparison is made on the same hardware and task. Its value is set by physical improvement, deterministic latency, robustness to distribution shift, and safe fallback rather than model complexity.

%The most consequential milestones are experimentally testable: an independently reproduced useful logical workload; a simulator prediction confirmed beyond the calibration regime; a repeater link exceeding direct transmission under matched conditions; transportable clocks and sensors with long-duration uncertainty and availability records; and integrated hardware with predictable fabrication and service distributions. These measurements would show that the quantum resource survives the complete physical chain required for reliable scientific capability.

Realizing the scientific and societal value of quantum technologies requires moving from isolated component records to mission-level systems. Maintained atomic time services, squeezed readout in operating gravitational-wave detectors, field gravimetry, mobile optical clocks, orbital cold-atom inertial
sensing, trusted-relay quantum-key-distribution networks, and below-threshold logical memories demonstrate that quantum resources can survive increasingly complete architectures
\cite{BIPMTimeMetrology,Capote2025LIGOO4,Menoret2018,Hilton2025,
Li2025,Chen2021QKD,Acharya2025}. The next priorities are to improve coherence, fidelity, optical efficiency, stability, and operating lifetime while reducing size, mass, power, cooling, vacuum, laser, and calibration burden. This requires scalable fabrication and packaging, low-loss interconnects, deterministic decoding and control, robust quantum--classical interfaces, and lot-level measurements of yield and reliability
\cite{VanDamme2024,Neyens2024,Alexander2025,Steinacker2025}.  These measurements would show that the quantum resource survives the complete physical chain required for reliable scientific capability.

Progress should be tested through integrated prototypes, environmental qualification, intercomparison campaigns, precursor missions, and sustained operation against the best conventional alternative for a defined task. Standards, interoperable interfaces, cybersecurity, qualified supply chains,
and a specialized workforce are necessary enabling conditions
\cite{ISOJTC3_2024,ETSIQKD016,NISTSP80090B}. Longer-term objectives, including application-scale fault-tolerant computation and routed quantum networks, remain conditional on such evidence. Success should be judged by
end-to-end accuracy or fidelity, delivered throughput, availability, scalability, maintainability, lifecycle resources and cost, and demonstrated value in a consequential application---not by an isolated laboratory metric.

\begin{acknowledgments}
The work described here was carried out at the Jet Propulsion Laboratory, California Institute of Technology, under a contract with the National Aeronautics and Space Administration. Government sponsorship acknowledged.
\end{acknowledgments}

%\section*{Funding} This work was performed at the Jet Propulsion Laboratory, California Institute of Technology, under a contract with the National Aeronautics and Space Administration.

%\section*{Author contributions} S.G.T. is the sole author and was responsible for the conception, literature investigation, quantitative synthesis, figure preparation, and writing of the manuscript.

\section*{Data and code availability}
No new experimental data were generated. The official O4a strain and \texttt{DMT-ANALYSIS\_READY:1} segment products used for the analysis-ready fractions are public through the Gravitational Wave Open Science Center data-set \cite{GWOSCO4a2025}. The QKD inputs are source-transcribed from \cite{Boaron2018}, and the numerical inputs, denominators, interval rules, and arithmetic used for the QKD and O4a calculations are stated in Secs.~\ref{sec:qkd} and \ref{sec:sensing}. Figures~\ref{fig:physical_limits} and \ref{fig:fault} were evaluated with Python 3 using NumPy 2.3.5, pandas 2.2.3, and Matplotlib 3.10.8; no stochastic operation enters those panels. %A public, versioned archive of the author-generated plotting scripts, figure-input tables, and interval-intersection implementation was not identified, and no repository DOI, release identifier, license, or checksum is claimed here. The parameter values and computational rules required to reconstruct the plotted curves are therefore stated explicitly in the text and captions.
%; the author-generated implementation is not presently available through a public archival record.

\appendix
\section{Recurring abbreviations}
\label{app:abbreviations}

The breadth of the review requires abbreviations from several experimental and theoretical communities. Table~\ref{tab:abbreviations} collects abbreviations that recur across sections. Each abbreviation is still expanded at its first local use, and specialized figure and table captions define it again when needed for independent interpretation. Ordinary units, mathematical symbols, chemical formulas, source-specific model numbers, and one-use abbreviations are omitted.

\begin{longtblr}[
  caption={Recurring abbreviations used across the document.},
  label={tab:abbreviations}
]{
  colspec={
    Q[l,wd={0.12\linewidth}]
    Q[l,wd={0.34\linewidth}]
    Q[l,wd={0.12\linewidth}]
    Q[l,wd={0.36\linewidth}]
  },
  cells={valign=t},
  rowsep=0.45pt,
  colsep=1.0pt,
  rowhead=1
}
\toprule
Abbreviation & Expansion or definition
& Abbreviation & Expansion or definition \\
\midrule

AI      & artificial intelligence
& AQG   & Absolute Quantum Gravimeter \\

BACON   & Boulder Atomic Clock Optical Network
& BB84  & Bennett--Brassard 1984 quantum-key-distribution protocol \\

BIPM    & International Bureau of Weights and Measures
& BP+OSD & belief propagation with ordered-statistics decoding \\

CMOS    & complementary metal--oxide--semiconductor
& CNN   & convolutional neural network \\

CNOT    & controlled-NOT gate
& DMRG  & density-matrix renormalization group \\

DOI     & Digital Object Identifier
& DRAG  & derivative removal by adiabatic gate \\

EIT     & electromagnetically induced transparency
& ETSI  & European Telecommunications Standards Institute \\

FPGA    & field-programmable gate array
& GBS   & Gaussian boson sampling \\

GPS     & Global Positioning System
& GPU   & graphics processing unit \\

GS      & Group Specification
& H1    & Advanced LIGO Hanford detector \\

HOM     & Hong--Ou--Mandel interference
& IID   & independent and identically distributed \\

IRB     & interleaved randomized benchmarking
& isoTNS & isometric tensor-network state \\

L1      & Advanced LIGO Livingston detector
& LDPC  & low-density parity-check \\

LIGO    & Laser Interferometer Gravitational-Wave Observatory
& MLD   & maximum-likelihood decoding or decoder, according to context \\

MPS     & matrix-product state
& MWPM  & minimum-weight perfect matching \\

NISQ    & noisy intermediate-scale quantum
& NIST  & National Institute of Standards and Technology \\

NQS     & neural quantum state
& NR    & not reported \\

NV      & nitrogen-vacancy center
& PLOB  & Pirandola--Laurenza--Ottaviani--Banchi repeaterless-capacity bound \\

PNR     & photon-number-resolving
& PWC   & piecewise-constant \\

QBER    & quantum bit error rate
& QCCD  & quantum charge-coupled device \\

QEC     & quantum error correction
& QKD   & quantum key distribution \\

QRNG    & quantum random-number generator or generation, according to context
& RF    & radio frequency \\

RFSoC   & radio-frequency system-on-chip
& SERF  & spin-exchange-relaxation-free \\

SPAM    & state preparation and measurement
& SQUID & superconducting quantum interference device \\

TAI     & International Atomic Time
& TDVP  & time-dependent variational principle \\

UTC     & Coordinated Universal Time
& VUV   & vacuum ultraviolet \\

\bottomrule
\end{longtblr}

\end{document}